\documentclass[11pt,a4paper]{article}

\usepackage{amsmath,amssymb,amscd}

\usepackage{colordvi}

\newcounter{mnotecount}[section]

\newcommand{\R}{\mathbb{R}}
\newcommand{\C}{\mathbb{C}}

\newcommand{\g}{\gamma}
\renewcommand{\a}{\alpha}

\newcommand{\Abar}{\overline{\cal A}}
\newcommand{\Atil}{\widetilde{\cal A}}

\newcommand{\Z}{\mathbb{Z}}

\newcommand{\h}{{\cal H}}
\newcommand{\scal}[2]{\langle #1| #2\rangle}
\newcommand{\cyl}{{\rm Cyl}}

\newcommand{\CQ}{{\rm Cyl}_{\rm Q}}

\newcommand{\CC}{{\cal C}}
\newcommand{\id}{{\rm id}}

\newcommand{\ot}{\otimes}

\newcommand{\Aut}{{\rm Aut}}

\newcommand{\suq}{{SU_q(2)}}

\newcommand{\tl}[1]{\tilde{#1}}

\DeclareMathOperator{\eee}{\!\text{$e$}}
\newcommand{\src}{{\sideset{^s}{}{\eee}}}
\newcommand{\tar}{{\sideset{^t}{}{\eee}}}
\newcommand{\cxi}{\check{\xi}}
\newcommand{\cg}{{\check{\gamma}}}

\newtheorem{thr}{Theorem}
\newtheorem{lm}[thr]{Lemma}
\newtheorem{df}{Definition}

\def\be#1\ee{\begin{equation}#1\end{equation}}

\numberwithin{equation}{section}
\numberwithin{thr}{section}
\numberwithin{chr}{section}
\numberwithin{df}{section}

\begin{document}
\title{Quantum group connections}
\date{October 16, 2008}
\author{Jerzy Lewandowski, Andrzej Oko{\l}\'ow}
\maketitle
\begin{center}
{\it  Institute of Theoretical Physics, Warsaw University\\ ul. Ho\.{z}a 69, 00-681 Warsaw, Poland\smallskip\\
lewand@fuw.edu.pl, oko@fuw.edu.pl}
\end{center}
\medskip

\begin{abstract}
The Ahtekar-Isham $C^*$-algebra known from Loop Quantum Gravity is the algebra of continuous functions on the space of (generalized) connections with a compact structure Lie group. The algebra can be constructed by some inductive techniques from the $C^*$-algebra of continuous functions on the group and a family of graphs embedded in the manifold underlying the connections. We generalize the latter construction replacing the commutative $C^*$-algebra of continuous  functions on the group by a non-commutative $C^*$-algebra defining a compact quantum group.
\end{abstract}

\section{Introduction}

The underlying algebra of Loop Quantum Gravity   is the Ashtekar-Isham algebra
\cite{ai}-\cite{aldiff},\cite{LQG}. This is a commutative, unital
$C^*$-algebra which consists of the so called cylindrical functions
 defined on the space of $SU(2)$-connections over a given
3-manifold $\Sigma$. This algebra is dual, in the
Gel'fand-Neimark sense, to a compactification of the space of
the $SU(2)$-connections.  The Ashtekar-Isham algebra has an equivalent
definition which does not involve the connections at all \cite{aldiff,alproj}.
It can be defined by gluing the algebras $\{C^{(0)}(SU(2))^{\g}\}$ assigned to
graphs $\{\gamma\}$ embedded in the manifold $\Sigma$. In this construction,
the natural partial ordering in the family of graphs is applied to define
suitable inductive family of the algebras.
The Ashtekar-Isham algebra  is useful in LQG, because
it admits the action of the diffeomorphisms of $\Sigma$, and,
moreover, a natural, invariant  state generated by the Haar measure
\cite{al-hoop}.
The corresponding Hilbert space serves as the kinematical
Hilbert space of LQG.  The generalization of the Ashtekar-Isham
algebra to an arbitrary compact group (still classical) is quite natural.
The goal of our current work is a generalization of the Ashtekar-Isham algebra
to a {\em compact quantum group}, in particular to the $\suq$ 
group.

Applications of quantum groups in the context of quantum
gravity or lattice Yang-Mills theory are known in the literature. For example, quantum group spin-networks play important
role in 2+1 quantum gravity \cite{qsn}, there is quantum group Yang-Mills
theory on lattice \cite{qgYM}. Another class of works presents constructions
of  spectral triples for spaces of connections by using the
inductive family approach \cite{str}. Our generalization
goes in a third direction.  We are not satisfied by assigning a
$C^*$-algebra to a single lattice or a graph as in the generalizations
of the lattice Maxwell theory or constructions of quantum group
spin-networks. The key difficulty we
address is admitting sufficiently large family of embedded graphs
and defining the gluing of the algebras in a way  consistent with the
possible overlappings and other relations between  the embedded graphs.
In Loop Quantum Gravity that consistency ensures the so called
"continuum limit" of the theory: despite of using lattices and graphs
the full theory is considered continuous rather than discreet.
We would like to maintain that continuum limit property
in our quantum group generalization.

We begin our paper with  introducing the ingredients
that will be used in the presented constructions: the set of embedded
graphs directed by a suitable relation, and a compact quantum group
(Section \ref{sec:ingred}.
We also recall  the graph construction of the classical group
Ashtekar-Isham algebra, whose generalization is the goal of this
work (Section \ref{subsec:AI}).

Our first result is generalization of the definition of the
Ashtekar-Isham algebra to an arbitrary compact quantum group
(Sec. \ref{sec:connection}).
We name the obtained algebra a {\it quantum group connection space}.
The non-commutativity of the quantum group leads to a new element: we
need to endow a given quantum group with a
*-algebra isomorphism which is an involution  and anti-comultiplicative.
The isomorphism corresponds to switching  the orientation in a graph.
We call it {\it internal
framing}.\footnote{One could say, that this is the price payed for
non-introducing in $\Sigma$ any framing of edges of the graphs, as
it is usually done in the definition of the quantum group
spin-networks.}.

Next, we  characterize the set of the internal framings
(Section \ref{subsec:intfram}). We also  study
the dependence of a  quantum group connection space on the internal
framing and formulate
conditions upon which two different internal framings  lead to
isomorphic quantum group connection spaces (Section \ref{subsec:isoequi}).
Still, however, we do not show the existence of an internal framing
in a general compact quantum group.

Finally, we focus   on the $\suq$  connection spaces (Section
\ref{sec:suqconnections}). We find all the internal framings in $\suq$.
They form a family
which has a natural structure of the circle. Every two internal framings
are conjugate to each other by the action of the automorphism group of $\suq$.
Therefore, all the $\suq$ connection spaces (corresponding to different
framings) are isomorphic to each other.

Finally, we
formulate yet another, equivalent up to a $C^*$-algebra isomorphism,
definition of a $\suq$  connection space which does not
distinguish any of the internal framings (Sec. \ref{sec:fraimind}). The
definition democratically uses all the internal framings of the
$\suq$  quantum group.

\section{Ingredients}\label{sec:ingred}
The  ingredients we will use in the next section to define a quantum
group connection space are:
\begin{enumerate}
\item  The directed set Gra of embedded graphs in a (semi)analytic manifold $\Sigma$ and
\item A compact quantum group $(\CC,\Phi)$ and an internal
framing $\xi:\CC\rightarrow\CC$.
\end{enumerate}
The first  ingredient is  known in LQG \cite{LQG} . We will
introduce it in detail in this section for the sake of completeness.

The quantum group theory is well known to everybody. However, taking into account the
diversity of definitions of a quantum group, we will
outline below the Woronowicz's approach \cite{slw} applied in our work. The notion
of "internal framing" is new, it is born in the current paper. We give our
definition already in this section. Only later, however,
in Section \ref{subsec:indfam}, the internal framings  will emerge  as solutions to the consistency conditions necessary  for our quantum group generalization
of the definition of the Ashtekar-Isham algebra.

\subsection{Directed set of embedded graphs}\label{subsec:graphs}
Let $\Sigma$ be a  manifold of the differentiability class
C$^{(\omega)}$ (analytic)  or
C$^{({\rm s}\omega)}$ (semianalytic) \cite{lost}.
The semianalytic manifolds are less known. Heuristically they could
be described as "piecewise analytic". A  careful definition
was introduced by using the theory of semianalytic sets.
But the definitions formulated
in this subsection and in this paper are unsensitive of the difference
between "analytic" and "semianalytic".  Therefore, the reader unfamiliar
with semi-analytic manifolds can assume the analyticity.

\begin{df}
$(i)$ An  edge in $\Sigma$ is an oriented 1-dimensional
submanifold  with 2-element boundary. The points of the boundary will be
referred to as the end points.

$(ii)$ An  embedded graph in $\Sigma$ is a
finite set of  edges in $\Sigma$, such that  every two
distinct edges can share only one or the both endpoints.
 Given a graph embedded in $\Sigma$, its
elements are referred to as the edges of the graph $\gamma$, and their
endpoints --- the vertices of $\gamma$. \label{graph}
\end{df}

The set of all graphs  (edges) in the manifold $\Sigma$ will be denoted by
Gra ($\cal E$).

Gra admits a natural structure of a directed set. In order to introduce
the suitable "inequivalence" relation $\geq$ in Gra, we need more
terminology.

Given an edge $e$ in $\Sigma$, the symbol $e^{-1}$ will denote an edge obtained
form an edge $e$ by the change of its orientation.

Given an  edge $e$, its orientation allows us to distinguish
between endpoints of the edge --- one of the endpoints will be
called the {\em source} of the edge and will be denoted by $\src$,
and the other one will be called the {\em target} and denoted by
$\tar$.

Given two edges $e_1$, and $e_2$ in $\Sigma$ such that:
\begin{itemize}
\item $\src_1=\tar_2 = e_1\cap e_2$, and
\item $e_1\cup e_2$ is again an edge in $\Sigma$
\end{itemize}
we denote
$$e_1\circ e_2 \ :=\ e_1\cup e_2$$
and call that operation $(e_1,e_2)\mapsto e_1\circ e_2$ {\it composition}
of the edges.

Briefly, two embedded graphs $\gamma$ and $\gamma'$ are in the
relation $\gamma'\geq \gamma$ iff  every edge of $\gamma$ can be expressed
by the edges of $\gamma'$ by using the composition
and/or changing orientation. A precise definition of the relation $\geq$
in Gra reads

\begin{df} For every $\gamma,\gamma'\in{\rm Gra}$,
$$\gamma\ \geq \gamma' $$
iff for every $e\in\gamma$ either of the following conditions is true:
$(i)$ $e\in\gamma'$, or $(ii)$ $e^{-1}\in\gamma'$, or $(iii)$ there are
$e_1,\ldots,e_K\in\gamma'$ and $\kappa_1,\ldots\kappa_K\in\{1,-1\}$ such that $e=e_1^{\kappa_1}\circ\ldots\circ e_K^{\kappa_K}$.
\end{df}

The pair $({\rm Gra}, \geq)$ is preserved by the group of the diffeomorphisms
of $\Sigma$ (analytic or, respectively, semianalytic).

The key property of the relation $\geq$ in Gra is:
\medskip

\noindent{\bf Completeness of (Gra, $\geq$) \cite{al-hoop}:}
{\it For every} $\gamma,\gamma'\in {\rm Gra}$, {\it there is}
$\gamma''\in {\rm Gra}$
{\it such that}
\begin{equation}
\gamma''\ \geq \gamma.
\label{compl}
\end{equation}

\medskip

Hence the pair $({\rm Gra}, \geq)$ is a directed set, and the structure
is invariant with respect to the group of the (semi)analytic diffeomorphisms
of the manifold $\Sigma$.

The completeness is the result we wanted to ensure by assuming
the (semi)analyticity. For example, if all the smooth edges were allowed,
then the property (\ref{compl}) would break down.


\subsection{Compact quantum group and internal framing}\label{subsec:qgroup}

In the present subsection we review the basic notions concerning compact
quantum groups according to Woronowicz's theory described in  \cite{slw}.

A compact quantum  group $(\CC,\Phi)$ consists of a unital (separable) $C^*$-algebra $\CC$ and a unital $C^*$-algebra homomorphism $\Phi:\CC\rightarrow\CC\otimes\CC$ such that:
\begin{equation}
(\Phi\otimes\id)\Phi=(\id\otimes\Phi)\Phi
\label{phi}
\end{equation}
and the sets
\begin{gather}
\{ \ (a\otimes I) \Phi(b) \ | \ a,b\in\CC\ \}\label{a}\\
\{ \ (I\otimes a) \Phi(b) \ | \ a,b\in\CC\ \}\label{b}
\end{gather}
(where $I$ is the unit of $\CC$) are linearly dense subsets of $\CC\otimes\CC$.

{Consider a linear map $\zeta:\CC\rightarrow\CC$. We say that $\zeta$
is comultiplicative (anticomultiplicative) if and only if for every $a\in\CC$ it satisfies the first (the second) of the following conditions:
\begin{gather}
\Phi(\zeta(a))=(\zeta\otimes\zeta)\Phi(a) \label{comul},\\
\Phi(\zeta(a))=F(\zeta\otimes\zeta)\Phi(a) \label{anticomul},
\end{gather}
where given two algebras  $\CC_1$ and $\CC_2$, the map
$F:\CC_1\ot\CC_2\rightarrow\CC_2\ot\CC_1$ is the flip map defined as
$$F(a_1\ot a_2):=a_2\ot a_1.$$}

{We will distinguish between automorphisms of the $C^*$-algebra $\CC$
and {\em comultiplicative} automorphisms of the algebra by calling the latter
ones automorphisms of a {\em quantum group} $(\CC,\Phi)$
(for short: automorphism of a quantum group $\CC$).}

{We are in a position now, to introduce  a new definition:}

{\begin{df}\label{framing}
An internal framing of a quantum group $(\CC,\Phi)$ is an
anticomultiplicative $C^*$-algebra automorphism
$$\xi:\CC\ \rightarrow \CC$$
such that
$$ \xi^2\ =\ {\rm id}.$$
\end{df}
}

Given two compact quantum groups $(\CC_1,\Phi_1)$ and $(\CC_2,\Phi_2)$,
the tensor product
$$ \CC_{12}:=\CC_1\ot\CC_2 $$
also admits a natural compact quantum group structure
$(\CC_{12},\Phi_{12})$. Indeed, define
$$\Phi_{12}:\CC_{12}\rightarrow\CC_{12}\ot\CC_{12}$$
to be  such that
\[
\Phi_{12}(a_1\ot a_2)=(\id\ot F\ot\id)(\Phi_1(a_1)\ot\Phi_2(a_2)).
\]
The pair $(\CC_{12},\Phi_{12})$
is a compact quantum group called the tensor product of the compact
quantum groups $(\CC_1,\Phi_1)$ and $(\CC_2,\Phi_2)$. Given compact
quantum groups $(\CC_i,\Phi_i)$ $(i=1,2,3)$, we can define a compact quantum group structure on $\CC_1\ot\CC_2\ot\CC_3$ by defining the structure first on $\CC_1\ot\CC_2$ (or on $\CC_2\ot\CC_3$) and then on $\CC_{12}\ot\CC_3$ (or on $\CC_1\ot\CC_{23}$). One can easily show that the both resulting structures on $\CC_1\ot\CC_2\ot\CC_3$ are isomorphic. Thus the notion of a tensor product of finitely many compact quantum groups is naturally defined.

\section{Quantum group connection space \label{sec:connection}}

This section is devoted to the basic object under our interest, that is,
to a  quantum group  generalization of the Ashtekar-Isham  algebra
defined by a compact topological group.

\subsection{The Ashtekar-Isham algebra}\label{subsec:AI}

We  start  with a brief
description of the (classical group) Ashtekar-Isham algebra (the description follows
\cite{proj,aldiff}).
The algebra is associated to  a (semi)analytic
manifold $\Sigma$  and a compact group $G$.

To every edge $e$ in $\Sigma$ we assign
\begin{equation}
C^e\ :=\ C^0(G^e),
\end{equation}
where $G^e$ is a set of all maps from $\{e\}$ to $G$ equipped with a topology induced by a natural bijection from $G^e$ onto $G$. Next, to each embedded graph (see Section \ref{subsec:graphs})  $\gamma$
we assign the $C^*$-algebra
\begin{equation}
C^\g\ :=\ C^0(G^\gamma)
\end{equation}
where $G^\gamma$ is the set from $\g=\{e_1,...,e_N\}$ to $G$ equipped with a topology induced by a natural bijection from $G^\g$ onto $G^N$. Clearly, $G^\g\cong G^{e_1}\times\ldots\times G^{e_N}$, hence (by virtue of Stone-Weierstrass theorem)
\begin{equation}
C^\g\cong C^{e_1}\ot\ldots\ot C^{e_N}.
\label{CCC}
\end{equation}

 Given a pair of graphs such that $\g'\geq\g$, there is a naturally defined injective unital $*$-homomorphism
\begin{equation}p_{\g'\g}:C^\g\rightarrow C^{\g'}.\label{pg'g}
\end{equation}
The family of the homomorphisms $(p_{\g\g'})_{\g'\geq\g\in{\rm Gra}}$
is consistent with the relation "$\geq$" in the set Gra in the following sense: for any triple of graphs such that $\g''\geq\g'\geq\g$, the corresponding maps satisfy
\begin{equation} 
p_{\g''\g}=p_{\g''\g'}\circ p_{\g'\g}
\label{pcons}.
\end{equation}
All the homomorphisms (\ref{pg'g}) can be determined by the consistency
(\ref{pcons})
and by fixing their action  in three elementary cases.

The first elementary case is the graph $\g'$  given by splitting an edge
of a 1-edge graph $\gamma$,
$$\g\ =\ \{e=e'\circ e''\},\ \ \ \g'\ =\ \{e',e''\} .$$
Then,
\begin{equation}\label{classsub}
p_{\g'\g}(f)(g(e'),g(e''))\ :=\ f(g(e')g(e'')).
\end{equation}

The second elementary case is  the graph $\g'$  given by switching  the orientation
in an edge $e\in\gamma$,
$$\g\ =\ \{e\},\ \ \ \ \g'\ =\ \{e^{-1}\}.$$
Now,
\begin{equation}\label{classinv}
p_{\g'\g}(f)(g(e^{-1}))\ :=\ f(g(e^{-1})^{-1}).
\end{equation}

The third elementary case is the graph $\g$ obtained from a graph $\gamma'$
by removing an edge,
$$\g\ =\ \{e_1,...,e_n\},\ \ \ \ \ \ \g'\ = \{e_1,...,e_n,e_{n+1}\}. $$
Here,
\begin{equation}
\label{classadd}
p_{\g'\g}(f)(g(e_1),...,g(e_n),g(e_{n+1}))\ :=\ f(g(e_1),...,g(e_n)).
\end{equation}

The resulting family $(C^\g,p_{\g'\g})_{\g\in{\rm Gra}}$ labeled by graphs in $\rm Gra$ is an inductive family of $C^*$-algebras. 
The inductive limit of the family  $(C^\g,p_{\g'\g})$ has a natural structure of a $C^*$-algebra. This is the Ashtekar-Isham algebra \cite{ai}.

Our aim, in this section, is to generalize that construction from a compact group $G$ to a compact quantum group. The difficulty will be in the non-commutativity. We will achieve the goal  in three steps: $(i)$ first, given  graph $\g$ embedded in $\Sigma$, we will associate with it a  (non-commutative in general) $C^*$-algebra $\CC^\g$, $(ii)$ then we will define appropriate $*$-homomorphisms $p_{\g'\g}:\CC^\g\rightarrow\CC^{\g'}$ obtaining an inductive family of $C^*$-algebras, $(iii)$ finally we will define the desired algebra as the inductive limit of the family.

\subsection{Algebras assigned to the graphs}\label{subsec:CCgamma}

Given a compact quantum group $(\CC,\Phi)$, assign with every edge $e$ in $\Sigma$ a $C^*$-algebra $\CC^{e}$ being a copy of $\CC$.

Consider an embedded graph
$$\g=\{e_1,\ldots,e_N\}.$$
We are going to define a $C^*$-algebra $\CC^\g$ associated with the
graph $\g$ as a tensor product of the algebras $\{\CC^{e_I}\}$ (see \eqref{CCC}). However,
the tensor product $\CC^{e_1}\otimes\dots\otimes\CC^{e_N}$ is ordering dependent,
while in general there is no natural way of ordering  edges of a graph.
Therefore, we use the ordering independent tensor product\footnote{This
tensor product can be defined by introducing  in the space of maps
${\rm Map}(\gamma,\CC)$ the suitable equivalence
relations. The definition we use is equivalent.}
\begin{equation}
\CC^\gamma\ =\ \bigotimes_{e\in\g}\CC^e.
\end{equation}
It can be introduced in the following way.

Choose any ordering $\sigma=(e_1,...,e_N)$ in the graph
$\g=\{e_1,...,e_N\}$ and assign to $\sigma$ the  tensor product
of the algebras:
\[
(e_1,...,e_N)\mapsto \CC^{e_{1}}\ot\ldots\ot\CC^{e_N}=:\CC^\gamma_\sigma.
\]
Let $\boldsymbol{\sigma}$ be a set of all the orderings.
For $\sigma,\sigma'\in\boldsymbol{\sigma}$ there is a natural quantum group
isomorphism\footnote{Given two quantum groups $(\CC_1,\Phi_1)$ and $(\CC_2,\Phi_2)$ a map $\rho:\CC_1\to\CC_2$ is a quantum group isomorphism if $(\rho\ot\rho)\circ\Phi_1=\Phi_2\circ\rho$.} $S_{\sigma,\sigma'}:\CC^\gamma_\sigma\rightarrow\CC^\gamma_{\sigma'}$,
\[
\CC^\gamma_\sigma\ni \ a_1\ot\ldots \ot a_N \ \mapsto \ a_{\sigma'\circ\sigma^{-1}(1)}\ot\ldots \ot a_{\sigma'\circ\sigma^{-1}(N)}\ \in\CC^\gamma_{\sigma'}.
\]
Now we consider the disjoint union
$$ \bigsqcup_{\sigma}\CC^\gamma_\sigma $$
and therein define the following equivalence relation:
given $a\in\CC^\g_\sigma$ and $a'\in\CC^\g_{\sigma'}$,
\begin{equation}
a \sim a' \ \ \text{iff} \ \ S_{\sigma,\sigma'}(a)=a'.
\label{rel-perm}
\end{equation}
Finally,
\[
\CC^\gamma:=\big(\bigsqcup_{\sigma\in\boldsymbol{\sigma}}\CC^\gamma_\sigma\big)/\sim.
\]
Because all maps $S_{\sigma,\sigma'}$ are isomorphisms of appropriate
quantum groups then  there exists a natural structure of compact
quantum group on $\CC^\gamma$. In the sequel we will assume that an ordering $\sigma$ of the edges of $\g$ is fixed and will work with the corresponding algebra $\CC^\g_\sigma$ as a representative of $\CC^\g$.

\subsection{Inductive family of $C^*$-algebras \label{subsec:indfam}}

The goal of this subsection is generalization of the maps $p_{\g'\g}$
defined in Section \ref{subsec:AI}, that is a construction of injective
unital
$*$-ho\-mo\-mor\-phisms $\{p_{\g'\g}:\CC^\g\rightarrow\CC^{\g'}\}$
defined for every pair $(\g',\g)$ of graphs such that $\g'\geq\g$.
As a result of this construction we will obtain the desired inductive
family of $C^*$-algebras.

Notice first that $\g\geq\g$. In this case we define $p_{\g\g}:=\id$.

If $\g'\geq\g$ and $\g'\neq\g$ then the graph $\g'$ can be obtained from
the graph $\g$ by means of the following elementary transformations:
\begin{enumerate}
\item ${\rm sub}_v$, subdividing an edge by adding a new vertex $v$,
\item ${\rm or}_e$, changing the orientation of an edge $e$,
\item ${\rm add}_e$, adding to the graph a new edge $e$.
\end{enumerate}
That is, there exists a finite sequence $(\g_i)$ $(i=1,\ldots,n)$  of
graphs such that $\g_1=\g$, $\g_n=\g'$, $\gamma_{i+1}\geq\g_i$  and
every graph $\g_{i+1}$ can be obtained from the graph $\g_i$ by precisely
one of the elementary transformations. To define $p_{\g'\g}$ we will define
first the family  of maps  $(p_{\g_{i+1}\g_{i}})$ corresponding to the
elementary transformations. Then,
\[
p_{\g'\g}:=p_{\g_n\g_{n-1}}\circ\ldots\circ p_{\g_2\g_1}.
\]
We emphasize, however, that a sequence of elementary transformations turning $\g$ into
$\g'$ is not uniquely defined. That will lead to consistency conditions we will
have to solve.

Generalizing the classical group case \eqref{classsub},\eqref{classinv},\eqref{classadd} let us assume that the homomorphisms corresponding to the elementary graph transformations have the following form:
\begin{enumerate}
\item subdividing an edge of $\g_i$. Let us order  the edges of
$\g_i$ in such a way that the subdivided edge is the last one, i.e. let
$\g_i=(e_1,\ldots,e_{N-1},e_N)$  and $\g_{i+1}=
(e_1,\ldots,e_{N-1},e'_N,e''_N)$ where $e_N=e'_N\circ e''_N$ and the ordering
of the last two edges in $\gamma_{j+1}$ is also relevant. Then
\begin{equation}
p_{\gamma_{i+1}\g_i}\ :=\ \id\ot\ldots\ot\id\ot\Phi\ =:\ p^{\rm sub},
\label{sub-phi}
\end{equation}
where $\Phi$ is the comultiplication of the quantum group, is a map from $\CC^{\g_i}$ into $\CC^{\g_{i+1}}$ corresponding to the transformation ${\rm sub}_v$ (provided $v$ is the new vertex subdividing $e_N$ onto $e'_N$ and $e''_N$).
\item changing the orientation. Assume that we change orientation of the edge $e_N$ obtaining an edge $e^{-1}_N$ i.e. $\g_i=(e_1,\ldots,e_{N-1},e_N)$ and $\g_{i+1}=(e_1,\ldots,e_{N-1},e^{-1}_N)$. Then
\begin{equation}
p_{\gamma_{i+1}\g_i}\ :=\ \id\ot\ldots\ot\id\ot\xi\ =:\ p^{\rm or},
\label{or-xi}
\end{equation}
where $\xi:\CC\rightarrow\CC$ is a *-homomorphism
(whose properties will be established below), is a map from $\CC^{\g_i}$
into $\CC^{\g_{i+1}}$ which corresponds to the transformation ${\rm or}_{e_N}$.
\item adding a new edge $e_{N+1}$. Let us fist define an inclusion
\[
\CC^{\ot^N}a\ni\mapsto{\rm In}(a)=a\ot I\in \CC^{\ot^{N+1}},
\]
\noindent where $I$ is the unit of $\CC$. Assuming the following ordering of the edges $\g_i=(e_1,\ldots,e_N)$ and $\g_{i+1}=(e_1,\ldots,e_{N},e_{N+1})$ the following map
\begin{equation}
p_{\gamma_{i+1}\g_i}\ :=\ {\rm In}\ = :\ p^{\rm add}.
\label{add-in}
\end{equation}
from $\CC^{\g_i}$ into $\CC^{\g_{i+1}}$ corresponds to the transformation ${\rm add}_{e_{N+1}}$.
\end{enumerate}

As we have already mentioned, the elementary graph transformations
sub$_v$, or$_e$, add$_e$ satisfy some identities. They imply
conditions on the maps  $p^{\rm sub}$, $p^{\rm or}$ and $p^{\rm add}$.
The identities are  the 'commutation relations' between the elementary
transformations. They read as follows (the vertex $v$ subdivides the edge $e$ into
edges $e_1$ and $e_2$, such that $e=e_1\circ e_2$):
\begin{align}
{\rm sub}_{v}\circ{\rm sub}_{v'}&={\rm sub}_{v'}\circ{\rm sub}_{v},\label{sub-sub}\\
{\rm sub}_{v}\circ{\rm add}_{e}&=
\begin{cases}
{\rm add}_{e}\circ{\rm sub}_{v}& \text{if $v\not\in e$}\\
{\rm add}_{e_1}\circ{\rm add}_{e_2}& \text{if $v\in e$}
\end{cases}\label{sub-add}\\
{\rm sub}_{v}\circ{\rm or}_e&=
\begin{cases}
{\rm or}_e\circ{\rm sub}_{v}& \text{if $v\not\in e$}\\
{\rm or}_{e_1}\circ{\rm or}_{e_2}\circ{\rm sub}_{v}& \text{if $v\in e$}
\end{cases},\label{sub-or}\\
{\rm or}_e\circ{\rm or}_{e'}&=
\begin{cases}
{\rm or}_{e'}\circ{\rm or}_e& \text{if $e\neq e'$}\\
\id& \text{if $e=e'$}
\end{cases},\label{or-or}\\
{\rm or}_e\circ{\rm add}_{e'}&=
\begin{cases}
{\rm add}_{e'}\circ{\rm or}_e& \text{if $e\neq e'$}\\
{\rm add}_{e^{-1}}& \text{if $e=e'$}
\end{cases}\label{or-add}\\
{\rm add}_{e}\circ{\rm add}_{e'}&={\rm add}_{e'}\circ{\rm add}_{e}\nonumber.
\end{align}

Let us analyze these of the above relations which impose restrictions on
the maps $\Phi$, $\xi$ and $\rm In$ applied to define the maps
$p^{\rm sub}$, $p^{\rm or}$ and $p^{\rm add}$.

If the new vertices $v,v'$ subdivide the same edge $e$   then the relation
\eqref{sub-sub} boils down to
\begin{equation}
(\Phi\otimes\id)\Phi=(\id\otimes\Phi)\Phi,
\end{equation}
that is the identity \eqref{phi} satisfied by the comultiplication
by the definition of a quantum group. The relation \eqref{sub-add} in the case $v\in e$ gives us
 $$\Phi(I)=I\ot I$$
which again is an identity for every compact quantum group. This means, that the definition of a quantum group is well suited for our
purpose!

  Notice now that the relation \eqref{sub-or} for $v\in e$ can be expressed
   as $e_2^{-1}\circ e_1^{-1}=(e_1\circ e_2)^{-1}$ which leads to the
   condition
   $$F\circ(\xi\ot\xi)\circ\Phi=\Phi\circ\xi$$ imposed on $\xi$.
   From \eqref{or-or}
  we get $\xi^2={\rm id}$ and the case $e=e'$ of \eqref{or-add} gives us
\begin{equation}\xi(I)=I.\label{idem}
\end{equation}
The latter condition is satisfied because we assumed $\xi$ to be multiplicative and the result $\xi^2={\rm id}$ just obtained means that $\xi$ is a bijection.   

Summarizing, we have obtained the following conditions that have to be
imposed on $\xi$:
\begin{equation}
\xi^2=\id, \ \ F\circ(\xi\ot\xi)\circ\Phi=\Phi\circ\xi.
\label{xi-inv}
\end{equation}
Hence, the condition is that $\xi$ be an internal framing in the quantum
group $(\CC,\Phi)$ (see Definition \ref{framing}).

Precisely, the following is true
\begin{lm}
Let $(\CC,\Phi)$ is a compact quantum group. Suppose
$$\xi:\CC\mapsto \CC$$
is an internal framing. Then, the formulae
(\ref{sub-phi},\ref{or-xi},\ref{add-in}) define a unique
family
$(p_{\g'\g})_{\g'\ge\g\in{\rm Gra}}$ of injective *-algebra
homomorphisms
$$p_{\g'\g}:\CC^{\gamma}\ \rightarrow\ \CC^{\gamma'} $$
such that
\begin{equation}
p_{\g''\g}\ =\ p_{\g''\g'}\circ p_{\g'\g}\label{p=pp},
\end{equation}
for every triple of embedded graphs such that $\g''\geq\g'\ge\g$.
\end{lm}

Thus, the family
\begin{equation}
(\CC^\g,p_{\g'\g})_{\g'\geq\g\in{\rm Gra}}
\label{ind}
\end{equation}
is an inductive family of $C^*$-algebras.

\subsection{The quantum group connection space}\label{subsec:qgc}

Let us construct the  limit of the inductive family
$(\CC^\g,p_{\g'\g})_{\g'\geq\g\in{\rm Gra}}$ and the C$^*$-algebra
structure induced therein by the compact quantum group structure
of $(\CC,\Phi)$.

In the disjoint union
$$ \bigsqcup_{\gamma\in {\rm Gra}}\CC^\gamma$$
we introduce the following equivalence relation:
given $a_{\gamma}\in\CC^\gamma$ and $b_{\gamma'}\in\CC^{\gamma'}$,
\begin{equation}
a_{\gamma}\sim b_{\gamma'}\  \ \text{iff}\ \  p_{\gamma''\gamma}\,a_{\gamma}\ =\ p_{\gamma''\gamma'}\,b_{\gamma'}
\label{rel-p}
\end{equation}
for every\footnote{Using \eqref{p=pp} one can easily show that if the
equality in the r.h.s. of \eqref{rel-p} is satisfied for a graph
$\g''_0\geq\g,\g'$, then it is satisfied for every graph $\g''\geq\g,\g'$.}
 $\gamma''\geq\gamma,\gamma'$. The quotient space 
\begin{equation}
\big( \bigsqcup_{\gamma\in{\rm Gra}} \CC^\gamma \big)/\sim
\label{quotient}
\end{equation}
can be equipped with the structure of a normed $*$-algebra\footnote{The quotient space is not a $C^*$-algebra yet since the space is not complete in the norm.}.
Indeed, denoting by $[a_\g]$ a general element of the quotient space (where
$a_\g\in\CC^\g$) one defines
\begin{gather*}
\lambda[a_\gamma]:=[\lambda a_\gamma],\ \ \ [a_\gamma]+[b_{\gamma'}]:=[\tilde{a}_{\gamma''}+\tilde{b}_{\gamma''}],\\
[a_\gamma][b_{\gamma'}]:=[\tilde{a}_{\gamma''}\tilde{b}_{\gamma''}],\ \ \ [a_\gamma]^*:=[a^*_\gamma],\\
\|[a_\gamma]\|:=\|a_\gamma\|,
\end{gather*}
where $a_\gamma\sim\tilde{a}_{\gamma''}$, $b_{\gamma'}\sim\tilde{b}_{\gamma''}$ and $\gamma''\geq\gamma,\gamma'$. The definitions of all operations and of the norm are consistent, because maps $\Phi$ and $\xi$ defining maps $p_{\g'\g}$ are injective (that is norm-preserving) $*$-homomorphism on $\CC$. Now, one completes the algebra \eqref{quotient} in the norm:

\begin{df}
The quantum group connection space defined by a (semi-)analytic manifold
$\Sigma$, a compact quantum group $(\CC,\Phi)$ and an internal framing $\xi$
is the inductive limit of the family $(\CC^\g,p_{\g,\g'})$,
\[
\CQ(\xi):=\underset{\longrightarrow}{\lim}\,\CC_\g:=\overline{\big( \bigsqcup_{\gamma\in{\rm Gra}} \CC_\gamma \big)/\sim}.
\]
equipped with the induced C$^*$-algebra structure.
\end{df}

We emphasize that, given a manifold $\Sigma$ and a compact quantum group
$(\CC,\Phi)$, the only additional element of the construction of the
algebra $\CQ(\xi)$ which is not given naturally, is the internal framing
$\xi$.
In fact, we can not guarantee the existence of such an internal framing
in every compact quantum group. However, we will be able to find all the
internal framings in the quantum group $\suq$.

Notice finally that, given a graph $\g$, there exists a natural injective $*$-homomorphism $p_\g$ from $\CC^\g$ into $\CQ$,
\begin{equation}
\CC^\g\ni a\mapsto p_\g(a):=[a]\in\CQ.
\label{inj-hom}
\end{equation}
Clearly, for $\g'\geq\g$ we have
\[
p_{\g'}=p_{\g'\g}\circ p_\g.
\]

\section{Dependence  on  internal framing}\label{sec:dependence}

Now,  one can ask whether two quantum group connection spaces
$\CQ(\xi)$ and $\CQ(\xi')$ built over the same quantum group and
base manifold can or cannot be isomorphic.  In this section
we formulate two different conditions, each of which is sufficient
for the existence of an isomorphism.

Even in the case of a classical
group, there are internal framings different then the canonical one dual to
$G\ni g\mapsto g^{-1}\in G$.
We discuss the dependence of the algebra $\cyl_Q(\xi)$  on $\xi$ in the classical
group case in the subsequent part of this section.
It turns out our quantum group generalization of the connections leads
also to a classical group generalization.

\subsection{Properties of the internal framings}\label{subsec:intfram}
Given a compact quantum group $(\CC,\Phi)$, denote by $\Xi(\CC,\Phi)$
the set of the internal framings, and let $\Aut(\CC,\Phi)$ stand for the group of all the automorphisms of the quantum group.

Let us study closer the relation between the internal framings and the quantum group automorphisms. The first observation is

\begin{lm} For every pair $\xi,\xi'\in \Xi(\CC,\Phi)$,
$$ \xi\circ \xi'\ \in\ \Aut(\CC,\Phi). $$
\label{xi-xi-aut}
\end{lm}
\noindent{\bf Proof.} Since $\xi,\xi'\in\Xi(\CC,\Phi)$ are
anticomultiplicative automorphisms of $\CC$, their composition is a
comultiplicative automorphism of $\CC$, that is, an automorphism of $(\CC,\Phi)$. $\blacksquare$

Using that observation, we can characterize
the set $\Xi(\CC,\Phi)$ of the internal framings by a subset
of $\Aut(\CC,\Phi)$ defined by this lemma:

\begin{lm}\label{rhoxirhoxi}
Let $\xi_0\in \Xi(\CC,\Phi)$. Suppose $\rho\in \Aut(\CC,\Phi)$ and
$$\rho\circ\xi_0\circ\rho\circ\xi_0\ =\ {\rm id}.$$
Then,
\begin{equation}\label{rhoxi0} \rho\circ \xi_0\ \in \Xi(\CC,\Phi).
\end{equation}
Every internal framing can be written in the form (\ref{rhoxi0}).
\end{lm}

Using this lemma, we will derive all the internal framings in $\suq$ 
in Section \ref{subsec:suqcon}.

On the other hand, every automorphism, can be used to produce
an internal framing from another internal framing in the following way

\begin{lm}
Let $\xi_0\in \Xi(\CC,\Phi)$ and $\rho\in \Aut(\CC,\Phi)$. Then,
$$ \rho^{-1}\circ\xi\circ\rho\ \in \ \Xi(\CC,\Phi).$$
\end{lm}

This leads to the following equivalence relation

\begin{df}\label{rel}
Two internal framings $\xi,\xi'\in\Xi(\CC,\Phi)$ are equivalent,
if and only if there is $\rho\in\Aut(\CC,\Phi)$ such that
$$\xi=\rho\circ\xi'\circ\rho^{-1}.
$$
The equivalence relation will be denoted by "$\sim$" and $\rho$
is called an intertwiner.
\end{df}

\subsection{Isomorphism from equivalence of internal framings}
\label{subsec:isoequi}

Given two equivalent internal framings,
an intertwiner is not, in general, uniquely defined. However each of
them gives rise to an isomorphism between the corresponding  quantum
group connection spaces:
\begin{lm}
Let $(\CC,\Phi)$ be a compact quantum group. Suppose two internal framings
$\xi,\xi'\in \Xi(\CC,\Phi)$ are equivalent and $\rho\in Aut(\CC,\Phi)$ is
their intertwiner. There is a  C$^*$-algebra isomorphism
$$\CQ(\xi)\ \rightarrow\ \CQ(\xi')$$
uniquely determined by $\rho$.
\label{iso-equi}
\end{lm}

\noindent {\bf Proof.} Algebras $\CQ(\xi)$ and $\CQ(\xi')$ are defined by
inductive families
\[
(\CC^\g,p_{\g'\g})_{\g\in{\rm Gra}}\ \ \text{and} \ \ (\CC^\g,p'_{\g'\g})_{\g\in{\rm Gra}}
\]
respectively, where the maps $p_{\g'\g}$ are defined by means of $\xi$  and
the maps $p'_{\g'\g}$ --- by means of $\xi'$.

 To show that the algebras are isomorphic it is enough to prove the existence of a family of maps
$(\omega_\g)_{\g\in{\rm Gra}}$ such that $(i)$ every $\omega_\gamma$ is
an  automorphism of the quantum group $\CC^\g$ and $(ii)$ for every pair
$\g'\geq\g$
\begin{equation}
 p'_{\g'\g}= \omega^{-1}_{\g'}\circ p_{\g'\g}\circ \omega_\g.
\label{omega-p}
\end{equation}
The corresponding isomorphisms is
$$ \cyl_Q(\xi)\ [a_\gamma]\ \mapsto\ [\omega_\gamma(a_\gamma)]\ \in \cyl_Q(\xi')  $$

To show that a given family of automorphisms $\{\omega_\g\}_{\g\in{\rm Gra}}$
satisfies (\ref{omega-p}) for every pair $\g'\geq\g$, it is necessary and sufficient
to show they satisfy the condition whenever $\g'$ is obtained from $\g$ by one of the three
elementary transformations.

Suppose that $\xi\sim\xi'$ in the sense of Definition \ref{rel} and $\rho$
is their intertwiner. Then, we define simply
\begin{equation}
\omega_\g:=\rho\ot\ldots\ot\rho.
\label{omega-nat}
\end{equation}
Since the maps $p_{\g'\g}$ are constructed from $\Phi$ and $\xi$, and
$\rho$ preserves the both structures, it is not hard to check, that
(\ref{omega-nat}) satisfies (\ref{omega-p}).
This completes the proof. $\blacksquare$

\subsection{Isomorphism from analytic structure in $\Sigma$}\label{subsec:isoanal}
It turns out that an analytic structure in the manifold $\Sigma$ can
be also used to construct an isomorphism between
two quantum group connection spaces $\cyl_Q(\xi)$ and $\cyl(\xi')$
defined by same manifold $\Sigma$ and quantum group $(\CC,\Phi)$ but
with two different internal framings $\xi$ and $\xi'$. We will present
the proof in this subsection. Now a clarification is in order. In our
paper, the manifold $\Sigma$ is allowed to have a weaker structure called
semianalytic, and the quantum group connection space may be constructed
using all the semianalytic graphs. The assumption we make in Lemma below is,
that the atlas of semianalytic charts contains a subatlas of analytic
charts. In the case of the quantum group connection spaces defined on analytic
manifolds, the assumption is trivially satisfied.
\begin{lm}\label{iso-analytic}
Let $(\CC,\Phi)$ be a compact quantum group and $\Sigma$ a semianalytic
manifold which admits an analytic atlas. Then, for every two internal framings
$\xi,\xi'$ in the quantum group, the corresponding quantum group connection spaces are isomorphic to each other.
\end{lm}

\noindent {\bf Proof.} Now let us restrict the family of graphs used so far: from the definition
of semianalytic edges it follows (see \cite{lost}) that every such an edge is a composition of a {\em finite} number of analytic edges, hence every
semianalytic graph can be transformed by adding new vertices (i.e. by one
of the three basic transformations) into a graph consisting of merely
{\em analytic} edges. Now, given an equivalence class
$[a_\gamma]\in\CQ(\xi)$, we can remove from it those elements
$\{a_{\g'}\}$ which are based on non-analytic graphs. The set of new
classes just obtained generates an algebra naturally isomorphic to
$\CQ(\xi)$---this new algebra will be identified with $\CQ(\xi)$
thereafter.

Since $\Sigma$ is now an analytic manifold we can consider a set
$\tilde{\cal E}^\omega$ of all analytically non-extendable analytic
oriented curves\footnote{For example the curve in $\Sigma=\R^2$ given
by equations $x(t)=t$ and $y(t)=\exp(-t^{-2})$ ($t\in]0,\infty[$) is
analytically non-extendable.} in $\Sigma$. Let $\tilde{o}:\tilde{\cal E}^\omega\rightarrow \mathbb{Z}_2=\{1,-1\}$ be a map such that
\[
\tilde{o}(\tilde{e})=-\tilde{o}(\tilde{e}^{-1})
\]
where $\tilde{e},\tilde{e}^{-1}$ are curves in $\tilde{\cal E}^\omega$ which occupy the same points in $\Sigma$ but have opposite orientations. Let ${\cal E}^\omega$ denotes the set of all analytic oriented edges. The map $\tilde{o}$ induces a map $o:{\cal E}^\omega\rightarrow\mathbb{Z}_2$:
\[
o(e):=\tilde{o}(\tilde{e}),
\]
where $e\subset\tilde{e}$  and the orientation of the edge coincides with the orientation of the curve (the definition is correct, because, given an edge $e$, there exists precisely one curve $\tilde{e}$ of the just described properties). Obviously,
\begin{equation}
o(e)=-o(e^{-1}).
\label{o}
\end{equation}

There exists (see Lemma \ref{xi-xi-aut}) $\rho\in\Aut$ such that $\xi'=\xi\circ\rho$. Then, given an analytic graph $\g$ with ordered edges $(e_1,\ldots,e_N)$, we define:
\[
\omega_\g:=\omega_{e_1}\ot\ldots\ot\omega_{e_N},
\]
where
\[
\omega_{e_I}:=
\begin{cases}
\id& \text{if $o(e_I)=1$}\\
\rho& \text{if $o(e_I)=-1$}
\end{cases}.
\]
Consider now two analytic graphs $\g'\geq\g$ such that (after some ordering of the edges) $\gamma=(e_1,\ldots,e_{N-1},e_N)$ and $\g'=(e_1,\ldots,e_{N-1},e^{-1}_N)$. Then $p_{\g'\g}$ is given by
\[
p^{\rm or}=\id\ot\ldots\ot\id\ot\xi
\]
and $p'_{\g'\g}$ --- by
\[
p^{\prime\rm or}=\id\ot\ldots\ot\id\ot\xi'.
\]
We have:
\begin{equation}
\omega^{-1}_{\g'}\circ p^{\rm or}\circ\omega_{\g}=\id\ot\ldots\ot\id\ot(\omega^{-1}_{e^{-1}_N}\circ\xi\circ\omega_{e_N}).
\label{omega-or}
\end{equation}
If $o(e_N)=1$ then
\[
\omega^{-1}_{e^{-1}_N}\circ\xi\circ\omega_{e_N}=\rho^{-1}\circ\xi\circ\id=\xi'
\]
(notice that $\xi'=\xi\circ\rho$ implies $\xi'=\rho^{-1}\circ\xi$). If $o(e_N)=-1$ then
\[
\omega^{-1}_{e^{-1}_N}\circ\xi\circ\omega_{e_N}=\id\circ\xi\circ\rho=\xi'.
\]
Thus (\ref{omega-or}) means that
\[
p^{\prime\rm or}=\omega^{-1}_{\g'}\circ p^{\rm or}\circ\omega_{\g}.
\]
Similarly we can show that in the case of appropriate graphs $\g',\g$
\[
p^{\prime\rm sub}=\omega^{-1}_{\g'}\circ p^{\rm sub}\circ\omega_{\g}\ \ \ \text{and}\ \ \ p^{\prime\rm add}=\omega^{-1}_{\g'}\circ p^{\rm add}\circ\omega_{\g}.
\]

The isomorphism $\Omega_{o}:\CQ(\xi')\rightarrow\CQ(\xi)$ given by a map $o:{\cal E}\mapsto \Z_2$ is a closure of the map:
\begin{equation}
\CQ(\xi')\ni[a_\g]\mapsto [ \omega_\g(a_\g) ] \in \CQ(\xi),
\label{Omega-iso}
\end{equation}
where $a_\g\in\CC^\g$ and $\g$ is analytic. $\blacksquare$

The isomorphism constructed in the proof depends essentially on
a choice of orientations of all the unextendable analytic curves in
$\Sigma$. Clearly,  that is a huge ambiguity.

\subsection{Discussion of the isomorphisms}\label{subsec:discussion} 

In the case of the existence of an intertwiner $\rho$ between two internal framings $\xi$ and $\xi'$, the isomorphism $\cyl_Q(\xi)\rightarrow\cyl(\xi')$ constructed in Lemma
\ref{iso-equi} is naturally defined. However, if $\rho'\not=\rho$ is
another intertwiner of the same pair of internal framings, than the isomorphism  defined by $\rho'$ is also different then the one defined by $\rho$.

The construction of an isomorphism $\cyl_Q(\xi)\rightarrow\cyl(\xi')$
from an analytic structure in $\Sigma$ presented in Lemma \ref{iso-analytic}
is quite general. In particular, it  does not use any assumption about the
internal framings. However each of those isomorphisms corresponds to
a choice of orientation of every unextendable analytic curve in $\Sigma$.
That huge ambiguity is the weakness of that result.
Another drawback is breaking of the diffeomorphism invariance.

Taking into account the above remarks it would be desirable to propose a framing independent definition of the quantum group connection space. This will be done  in the case of the quantum group $\suq$.

\subsection{Commutative case}\label{subsec:commutative}

Defining a  quantum group connection space as the $C^*$-algebra $\CQ$ we
did not assume that the $C^*$-algebra $\CC$ of a compact quantum group
is either commutative or non-commutative.  In particular we can take
the algebra $C^0(G)$ of a classical compact group $G$. We would like to study that case in this subsection.

Suppose that the algebra $\CC$ of a compact quantum group $(\CC,\Phi)$ is
commutative. Then $\CC=C^0(G)$, where $G$ is a (topological) compact group.

Let us apply Lemma \ref{rhoxirhoxi} and its  characterization of the
internal framings  to this case.

A canonically defined internal framing is
$$ \xi_0\ =\ \kappa$$
where $\kappa$ is the antipode map, that is the pullback to $C^0(G)$ of the map
$$\kappa_*: g\ \mapsto g^{-1}\in G.$$

Every $\rho\in \Aut(C^0(G),\Phi)$ is the pullback of a group automorphisms
$\rho_*\in\Aut(G)$. The condition
$$\rho\circ\kappa\circ\rho\circ\kappa \ =\ {\rm id}$$
is equivalent to
$$\kappa_*\circ\rho_*\circ\kappa_*\circ\rho_* \ =\ {\rm id}.$$
But since $\kappa_*$ is preserved by every group automorphism,
the latter condition is equivalent to
$$(\rho_*)^2\ =\ {\rm id}.$$

Hence,
\begin{lm}\label{rhokapparhokappa} Suppose that
$$\rho_*\in\Aut(G)\ \ \ {and}\ \ \ (\rho_*)^2\ =\ {\rm id}.$$
Then the pullback of the map
$$ \xi_*\ :=\ \kappa_*\circ \rho_*$$
is an internal framing of the commutative quantum group $(C^0(G),\Phi)$.
Every internal framing in $(C^0(G),\Phi)$ can be written in this form.
\end{lm}

Consider $\CQ(\xi)$ for some $\xi\in\Xi(C^0(G),\Phi)$ and denote by $\Atil$ the
Gel'\-fand spectrum of $\CQ(\xi)$.  The spectrum can be thought as a set of
maps $\tilde{A}:{\cal E}\rightarrow G$ such that:
\[
\tilde{A}(e_1\circ e_2)=\tilde{A}(e_1)\tilde{A}(e_2)
\]
and
\begin{equation}
\tilde{A}(e^{-1})=\xi_*(\tilde{A}(e)).
\label{xi*}
\end{equation}
The proof  is a simple generalization of the proof of Proposition 2 in
\cite{al-difgeom}.

The standard construction of the classical Ashtekar-Isham  algebra
based on a compact group $G$ uses
$$\xi_0=\kappa.$$

However, in general  there exist different framings
 $\xi\in\Xi(\CC)$ such that $\xi\neq\kappa$
 (it is the case if $\CC=SU_1(2)$). Notice,  that then
 $$\xi\not\sim\kappa$$
 in the sense of \eqref{rel}, because the antipode $\kappa$ is preserved
 by all the  automorphisms of $(C^0(G),\Phi)$.
Then the algebras $\CQ(\xi)$ and $ \CQ(\kappa)$ may not be isomorphic
if the manifold $\Sigma$ does not admit any analytic structure on it.
Otherwise an isomorphism $\Omega_o$ between the algebras can be  built by means of $\rho\in\Aut$ such that $\xi=\kappa\circ\rho$ and the map $(\Omega_o)_*:\Abar\rightarrow\Atil$ will be given by the following rule: $(\Omega_o)_*(\bar{A})$ is an element of $\Atil$ such that
\[
(\Omega_o)_*(\bar{A})(e)=
\begin{cases}
\bar{A}(e)& \text{if $o(e)=1$}\\
\rho_*(\bar{A}(e))& \text{if $o(e)=-1$}
\end{cases},
\]
where $\rho_*$ is an automorphism of the group $G$ induced by $\rho$.

\section{$\suq$ connection spaces}\label{sec:suqconnections}

Having described basic properties of the quantum group connection spaces,
we are going to study carefully the quantum group $SU_q(2)$ example.
As the reference we will be using \cite{slw-suq}. The choice of this
group is partially motivated by the fact that it is the simplest
non-trivial example of a compact quantum group. On the other hand a Lie
group $SU(2)$ is essential for many physical applications and therefore
it is interesting to study the 'quantum connections' understood as
deformations of $SU(2)$-connections.

Finally, we caution the reader that in the sequel we will neglect the
case $q=-1$, because the structure of  $SU_{-1}(2)$ differs
too much from the structure of the other groups $\suq$.

The task of constructing  quantum group connection spaces  boils down
to finding all the internal framings of
the quantum group $\suq$ and studying the equivalence relation.
We will separately consider two cases: $(i)$ the proper quantum group
$\suq$ with $q^2\neq1$, and the commutative $SU_1(2)$ group.
In particular we will find, that in the proper quantum group case,
all the internal framings are equivalent to each other.

\subsection{The quantum group $\suq$.}\label{subsec:suq}

Let us begin the study by recalling some basic facts concerning the
quantum group $\suq$ described in \cite{slw-suq, slw}.

The Hopf $*$-algebra $H_q$ of quantum group $\suq$ is generated by elements
\begin{equation}
\{\a,\a^*,\gamma,\gamma^*\}
\label{gen-0}
\end{equation}
satisfying the following relations:
\begin{equation}
\begin{gathered}
\a^*\a + \gamma^*\gamma=I,\ \ \ \a\a^*+q^2\gamma\gamma^*=I\\
\gamma^*\gamma=\gamma\gamma^*,\ \ q\g\a=\a\g, \ \ q\g^*\a=\a\g^*,
\end{gathered}
\label{suq}
\end{equation}
where $q\in[-1,1]\setminus\{0\}$.

The comultiplication $\Phi$ on $H_q$ is given by
\begin{equation}
\Phi(\a)=\a\ot\a-q^2\g^*\ot\g, \ \Phi(\g)=\g\ot\a+\a^*\ot\g.
\label{phi-su}
\end{equation}

Those relations are  encoded in the following matrix
\begin{equation}
(u_{ij})=
\begin{pmatrix}
\a & -q\gamma^*\\
\gamma & \a^*
\end{pmatrix}.
\label{su-fund}
\end{equation}
They read
\begin{equation}
\sum_k u^*_{ki}u_{kj}=I\delta_{ij}=\sum_k u_{ik}u^*_{jk},\ \ \ \ \ \Phi(u_{ij})\ =\ \sum_{k} u_{ik}\otimes u_{kj}.
\label{su-fund-rel}
\end{equation}

There is a naturally defined norm on $H_q$ which makes the
completion $\CC_q$ equipped with the extension of $\Phi$ a compact
quantum group $(\CC_q,\Phi)$ known as $\suq$.

Let us finally make  the following abbreviations:
\begin{equation}
\Aut(\suq,\Phi)\ =: \ \Aut_q,\ \ \ \ \ \ \Xi(\suq,\Phi)\ =:\ \Xi_q.
\end{equation}

\subsection{The  proper $\suq$ connection spaces}\label{subsec:suqcon}
Let us consider the quantum group $\suq$ for $q^2\neq 1$. We will construct all the integral framings by guessing\footnote{In fact, in the $\suq$ case with $q\neq -1$ it is possible to derive all internal framings from Definition \ref{framing}. This, however, requires an application of the representations theory of $\suq$, which makes the task more difficult than what we are going to present below.}
one of them, $\xi_0$, and, according to Lemma \ref{rhoxirhoxi}, solving the equation
$$ \rho\circ\xi_0\circ\rho\circ\xi_0\ =\ {\rm id} $$
for  automorphisms $\rho$ of $\suq$.

The group $\Aut_q$ of the automorphisms of the quantum group $\suq$ is isomorphic to $U(1)$ (see Appendix \eqref{ap-auth}). The isomorphism is defined by the following map
\[
U(1)\ni w \mapsto \rho_{w} \in \Aut_q,
\]
where according to \eqref{aut-q-1} 
\begin{equation}
\rho_{w}(\alpha)=\alpha, \ \ \ \rho_{w}(\gamma)=w\gamma.
\label{aut-a-g}
\end{equation}
An example of an internal framing $\xi_0\in\Xi_q$ is (see Appendix \ref{app-anti})
\begin{equation}
\xi_0(\alpha)\ =\ \alpha,\ \ \ \ \ \ \xi_0(\gamma)\ =\ \gamma^*.
\label{xi-ex}
\end{equation}
Now we can apply Lemma \ref{rhoxirhoxi}. By the direct calculation we find
that the condition
\begin{equation}
\rho_w\circ\xi_0\circ\rho_w\circ\xi_0(a)\ =\ a,\ \ \ \ \ a=\alpha,\gamma,
\end{equation}
is satisfied for every $\rho_{u}\in\Aut_q$. Hence,
\begin{equation}
\rho_{w}\circ\xi_0\circ\rho_{w}\circ\xi_0\ =\ {\rm id},
\end{equation}
is an identity true for every $\rho_w\in \Aut_q$. Therefore, the set $\Xi_q$ of the internal framings in $\suq$ can be characterized by the following bijective  map
\[
\Aut_q\ni\rho_{w}\mapsto  \rho_{w}\circ\xi_0\ :=\ \xi_{w}\in\Xi_q
\]
whose inverse is
\[
\Xi_q\ni\xi\mapsto\xi\circ\xi_0\in\Aut_q.
\]
The map endows $\Xi_q$ with  a topology (and a differential structure if needed) independent of the choice of $\xi_0$.  Indeed, if we chose a different  $\xi_1\in\Xi_q$, then
\[
\Xi_q\ni\xi\mapsto\xi\circ\xi_1\ = \xi\circ\xi_0\circ\xi_0\circ\xi_1=
\xi\circ\xi_0\circ\rho_{01}\in\Aut_q
\]
hence the difference in the mappings is the translation of $\Aut_q$ by an automorphism $\rho_{01}:=\xi_0\circ\xi_1$.

From \eqref{aut-a-g} and \eqref{xi-ex} we obtain the following
\begin{lm}
A general internal framing  of $\suq$ has the following form
\begin{equation}
\xi_{w}(\gamma)\ =\ \bar{w}\gamma^*,\ \ \ \ \ \
\xi_{w}(\alpha)\ =\ \alpha,\label{xiphi}
\end{equation}
where $w\bar{w}=1$. 
\end{lm}

Knowing explicitly  the general form of an internal framing we can
easily construct automorphisms of $\suq$ which map an internal framing
into another.

\begin{lm}
Let $\xi_{w},\xi_{w'}\in\Xi_q$ be defined by (\ref{xiphi}). Then, an automorphism $\rho_{z}\in\Aut_q$
satisfies
\begin{equation}
\rho_{z}\circ\xi_{w}\circ\rho^{-1}_{z}\ =\ \xi_{w'}
\end{equation}
if and only if
\begin{equation}
z^2=\bar{w}w'
\label{z2ww'}
\end{equation}
In particular, the automorphism $\rho_{\pm 1}\in\Aut_q$ preserves every
internal framing.
\end{lm}
\noindent{\bf Proof.} A calculation gives
\begin{equation}
\rho_{z}\circ\xi_{w}\circ\rho^{-1}_{z}(\gamma)\ =\ \bar{z^2}\bar{w}\gamma^*
\ =\ \xi_{z^2w}(\gamma).
\label{zzww}
\end{equation}
$\blacksquare$

Consequently, in the case of noncommutative $\suq$ there is
a 1-di\-men\-sio\-nal family (a circle) of the connection spaces
$\cyl_Q(\xi_{w})$ labeled by $w\in U(1)$. For every  pair
$\cyl_Q(\xi_{w})$ and  $\cyl_Q(\xi_{w'})$ there are two distinct $C^*$-algebra
 isomorphisms
$\cyl_Q(\xi_{w})\rightarrow\cyl_Q(\xi_{w'})$ given via Lemma \ref{iso-equi} by the $\suq$ automorphisms $\{\rho_{z},\rho_{-z}\}$ with $z$ being a solution of \eqref{z2ww'}. 

\subsection{The  $SU_1(2)$ quantum connection spaces }\label{su2con}

The $C^*$-algebra $\CC_1$ of the quantum group $SU_1(2)$ is commutative
and $SU_1(2)$ corresponds to the Lie group $SU(2)$. To build all possible
quantum $SU_1(2)$ connection spaces  we have to find the set $\Xi_1$  of
all internal framings  in $SU_1(2)$.  The latter ones  are given by
Lemma \ref{rhokapparhokappa} and all the involutive  automorphisms
of the group $SU(2)$. The automorphism group of $SU(2)$ is isomorphic
to the rotation group $SO(3)$. A non-trivial rotation $R_{X,\alpha}$ around an axis $X$ for the angle $\alpha$ is an involution if and only if $\alpha=\pi$ , and every non-trivial involutive element of $\Aut_1$ is a pullback $R^*_{X,\pi}$ of $R_{X,\pi}$ to $C^0(SU(2))$. Hence, every  internal framing  of $SU_1(2)$ is either the antipode  $\kappa$ or 
\begin{equation}
\xi_X =\ R^*_{X,\pi}\circ \kappa.
\label{Xpi}
\end{equation} 

Next,  recall  the equivalence relation \eqref{rel} defined on $\Xi_1$. It divides the set into two equivalence classes:
\begin{enumerate}
\item the first class consists of one element only---this is the antipode $\kappa$;
\item the second class consists of the internal framings given by (\ref{Xpi}). Indeed, given two axes $X$ and $X'$ there exists $R_{Y,\alpha}\in SO(3)$ such that 
\begin{equation}
R_{X,\pi}=R^{-1}_{Y,\alpha}\circ R_{X',\pi}\circ R_{Y,\alpha}.
\label{XYXY}
\end{equation} 
This together with \eqref{Xpi} gives 
\[
\xi_{X}=R^{*}_{Y,\alpha}\circ \xi_{X'}\circ R^{*-1}_{Y,\alpha}.      
\]
\end{enumerate}

For the completeness we translate those observation into the language of the quantum group description of $SU_1(2)$. The antipode acts on the quantum group generators as
\[
\kappa(\a)=\a^*, \ \ \ \kappa(\g)=-\g.
\]

The elements of the second equivalence class are
defined on the quantum group generators as follows (see Appendix \ref{app-anti})
\begin{equation}
\begin{aligned}
\xi_{X}(\a)&=z\bar{z}\a-z{x}\g+\bar{z}x\g^*+x^2\a^*,\\
\xi_{X}(\g)&=-\bar{z}{x}\a+\bar{x}^{ 2}\g+\bar{z}^{ 2}\g^*+\bar{z}{x}\a^*,
\end{aligned}
\label{xi-zx}
\end{equation}
where the axis $X$ is represented by a vector $({\rm Re}z,{\rm Im}z,x)\in\R^3$ such that $z\bar{z}+x^2=1$. We have $\xi_{X}=\xi_{X'}$ if and only if $X'=\pm X$.

We learn from that characterization that there are  two classes of algebras in the $SU_1(2)$ case. The first one contains only $\cyl_1(\kappa)$ which coincides with the Ashtekar-Isham algebra introduced in Section \ref{subsec:AI}. The second one consists of the algebras $\cyl_1(\xi_{X})$. According to the result of Section \ref{subsec:isoequi} each two algebras $\cyl_1(\xi_{X})$ and $\cyl_1(\xi_{X'})$ are always isomorphic, while $\cyl_1(\xi_{X})$ and $\cyl_1(\kappa)$ may not be so. It should be noted that there is no canonical isomorphism between $\cyl_1(\xi_{X})$ and $\cyl_1(\xi_{X'})$---an isomorphism between the algebras is constructed from an automorphism $\rho$ of $SU_1(2)$ such that $\xi_{X'}=\rho\circ\xi_{X}\circ\rho^{-1}$ and given $\xi_{X}$ and $\xi_{X'}$ there are a lot of such automorphisms (given $X,X'$ there are a lot of $(Y,\alpha)$ satisfying \eqref{XYXY}).

\section{A natural $\suq$ connection space}\label{sec:fraimind}

In this section we continue the study of the quantum group $\suq$ connection space. We turn now to an equivalent---up to the $C^*$-algebra isomor\-phisms---definition of the $\suq$ connection space introduced in a manner independent of choice of internal framing in $\suq$.

\subsection{Universal coverings of $\Aut_q$ and $\Xi_q$ }

Let $\check{\Aut}_q$ be the universal covering of $\Aut_q$. $\check{\Aut}_q$ is of infinite cardinality and is a Lie group isomorphic to the additive group $\R$  of real numbers. Let us denote by $\check{\Xi}_q$ the universal covering of $\Xi_q$. Clearly, $\check{\Xi}_q$ is homeomorphic to the real line $\R$.  We also denote by $\pi_{\rm A}$ and $\pi_\Xi$ the appropriate canonical projections:
\[
\pi_{\rm A}:\check{\Aut}_q\to\Aut_q, \ \ \ \ \pi_\Xi:\check{\Xi}_q\to\Xi_q.
\]
Consequently, we can parameterize the sets $\check{\Aut}_q$ and $\check{\Xi}_q$ by real numbers: given $\beta\in\R$, $\check{\rho}_\beta$ and $\cxi_\beta$ are elements of, respectively, $\check{\Aut}_q$ and $\check{\Xi}_q$ such that
\begin{equation}
\pi_{\rm A}(\check{\rho}_\beta)=\rho_{e^{i\beta}} \ \ \text{and} \ \ \pi_\Xi(\cxi_\beta)=\xi_{e^{i\beta}}.
\label{R-xi-aut}
\end{equation}

\subsection{Framed graphs and the corresponding algebras}
Let $\gamma$ be a graph consisting of edges $\{e_1,\ldots,e_N\}$. We associate with every edge of the graph an element of $\check{\Xi}_q$, $e_I\mapsto\check{\xi}_I$. Thus we obtain a {\em framed} graph $\check{\gamma}$ as the collection
\[
\check{\gamma}:=\{(e_1,\check{\xi}_1),\ldots,(e_N,\check{\xi}_N)\}.
\]
Since now the graph $\gamma$ will be called the {\em bare} graph of the framed graph $\check{\gamma}$. We define a directing relation on the set $\check{\rm Gra}$ of all framed graphs as
\[\label{geq}
\check{\g}'\geq\check{\g} \iff {\g}'\geq{\g},
\]
where the latter relation is the standard directing relation on the set of bare graphs. With every framed graph $\check{\g}$ we associate a noncommutative $C^*$-algebra
\[
\CC^{\check{\g}}:=\CC^\g\cong\CC_q^N,
\]
where $N$ is the number of edges of $\g$. Now, for every pair $\check{\g}'\geq\check{\g}$ we are going to define an injective homomorphism $p_{\check{\g}'\check{\g}}:\CC^{\check{\g}}\to\CC^{\check{\g}'}$.


Similarly to the case of bare graphs, if $\check{\g}'\geq\check{\g}$ then
the graph $\check{\g}'$ can be obtained from $\check{\g}$ by a sequence of the following elementary transformations:
\begin{enumerate}
\item ${\rm sub}_v$, subdividing an edge by adding a new vertex $v$ and equipping the two new edges with the framing associated with the original one;
\item ${\rm or}_e$, changing the orientation of an edge $e$ while preserving the framing;
\item ${\rm add}_{e,\cxi}$, adding to the graph a new edge $e$ with a framing $\cxi$;
\item ${\rm fr}_{e,\check{\xi}'\check{\xi}}$, changing the framing of an edge $e$ from $\check{\xi}$ to $\check{\xi}'$.
\end{enumerate}

\subsection{An inductive family}

As before, given $\check{\g}'\geq\check{\g}$, we will define $p_{\check{\g}'\check{\g}}$ as a composition of maps $p^{\rm sub},p^{\rm or},p^{\rm add}$ and $p^{\rm fr}$ corresponding respectively to the new elementary tran\-sformations. In the case of the first three maps $p^{\rm sub},p^{\rm or}$ and $p^{\rm add}$ we apply previous formulas \eqref{sub-phi}, \eqref{or-xi} and \eqref{add-in} respectively. To define the map $p^{\rm fr}$ assume that the graph $\cg$ possess $N$ edges and that we are going to change the framing of the $N$-th edge from $\check{\xi}$ to $\check{\xi}'$ obtaining as the result a framed graph $\cg'$. Then $p^{\rm fr}:\CC^\cg\to\CC^{\cg'}$ should be of the following form
\begin{equation}
p^{\rm fr}:=\id\ot\ldots\ot\id\ot f_{\cxi'\cxi},
\label{fr-f}
\end{equation}
where $f_{\cxi'\cxi}$ is an isomorphism of the $C^*$-algebra $\CC_q$.

Obviously, while transforming the graph $\cg$ to $\cg'$  the elementary transformations can be applied in different order. This means that we again have to consider 'commutation relations' between the transformations. 

\subsubsection{Commutation relations}

The 'commutation relations' between the first three new transformations resemble closely the previous ones (see Section \ref{subsec:indfam})---assuming that a vertex $v$ subdivide an edge $e$ into edges $e_1$ and $e_2$ such that $e=e_1\circ e_2$ we obtain:
\begin{equation}
\begin{aligned}
{\rm sub}_{v}\circ{\rm sub}_{v'}&={\rm sub}_{v'}\circ{\rm sub}_{v},\\
{\rm sub}_{v}\circ{\rm add}_{e,\cxi}&=
\begin{cases}
{\rm add}_{e,\cxi}\circ{\rm sub}_{v}& \text{if $v\not\in e$}\\
{\rm add}_{e_1,\cxi}\circ{\rm add}_{e_2,\cxi}& \text{if $v\in e$}
\end{cases}\\
{\rm sub}_{v}\circ{\rm or}_e&=
\begin{cases}
{\rm or}_e\circ{\rm sub}_{v}& \text{if $v\not\in e$}\\
{\rm or}_{e_1}\circ{\rm or}_{e_2}\circ{\rm sub}_{v}& \text{if $v\in e$}
\end{cases},\\
{\rm or}_e\circ{\rm or}_{e'}&=
\begin{cases}
{\rm or}_{e'}\circ{\rm or}_e& \text{if $e\neq e'$}\\
\id& \text{if $e=e'$}
\end{cases},\\
{\rm or}_e\circ{\rm add}_{e',\cxi}&=
\begin{cases}
{\rm add}_{e',\cxi}\circ{\rm or}_e& \text{if $e\neq e'$}\\
{\rm add}_{e^{-1},\cxi}& \text{if $e=e'$}
\end{cases}\\
{\rm add}_{e,\cxi}\circ{\rm add}_{e',\cxi'}&={\rm add}_{e',\cxi'}\circ{\rm add}_{e,\cxi}.
\end{aligned}
\label{three-trans}
\end{equation}

Consequently, $p^{\rm sub}$ is given as before by the map $\Phi$ defining the quantum group structure on $\CC_q$, while $p^{\rm or}$---by an element of $\Xi_q$. Given $p^{\rm or}$ acting on the copy of algebra $\CC_q$ associated with the framed edge $(e,\cxi)$ it is natural to require that $\xi$ defining $p^{\rm or}$ is given by
\[
\xi=\pi_{\Xi}(\cxi).
\]
Thus in the framework of the framed graphs the maps $p^{\rm sub},p^{\rm or}$ and $p^{\rm add}$ are naturally defined.

The `commutation relations' between the first three transformations and the fourth one are as follows:
\begin{align}
{\rm sub}_{v}\circ{\rm fr}_{e,\check{\xi}'\check{\xi}}&=
\begin{cases}
{\rm fr}_{e,\check{\xi}'\check{\xi}}\circ{\rm sub}_{v}& \text{if $v\not\in e$}\\{\rm fr}_{e_1,\check{\xi}'\check{\xi}}\circ{\rm fr}_{e_2,\check{\xi}\check{\xi}'}\circ{\rm sub}_{v}& \text{if $v\in e$}
\end{cases},\label{sub-fr}\\
{\rm or}_{e'}\circ{\rm fr}_{e,\check{\xi}'\check{\xi}}&=
{\rm fr}_{e,\check{\xi}'\check{\xi}}\circ{\rm or}_{e'},\label{or-fr}\\
{\rm fr}_{e,\check{\xi}''\check{\xi}}\circ{\rm add}_{e',\cxi'}&=
\begin{cases}
{\rm add}_{e',\cxi'}\circ{\rm fr}_{e,\check{\xi}''\check{\xi}'}& \text{if $e\neq e'$ }\\
{\rm add}_{e,\cxi''}& \text{if $e=e'$ and $\cxi=\cxi'$}
\end{cases},\label{add-fr}\\
{\rm fr}_{e,\cxi''\cxi'}\circ{\rm fr}_{e',\cxi'''\cxi}&=
\begin{cases}
{\rm fr}_{e',\cxi'''\cxi}\circ{\rm fr}_{e,\cxi''\cxi'}& \text{if $e\neq e'$}\\
{\rm fr}_{e,\cxi''\cxi} & \text{if $e=e'$ and $\cxi'=\cxi'''$}
\end{cases}\label{fr-fr}.
\end{align}
These relations impose some restrictions on the map $f_{\cxi'\cxi}$ defining $p^{\rm fr}$ via \eqref{fr-f}. Once an $f_{\cxi'\cxi}$ consistent with the restrictions is found the construction of the desired maps $\{p_{\cg'\cg}\}$ becomes straightforward.

\subsubsection{Solving consistency conditions imposed on $f_{\cxi'\cxi}$ }

The 'commutation relation' \eqref{sub-fr} implies that
\[
(f_{\cxi'\cxi}\ot f_{\cxi'\cxi})\circ\Phi=\Phi\circ f_{\cxi'\cxi},
\]
which means that $f_{\cxi'\cxi}\in\Aut_q$. The next relation, that is \eqref{or-fr}, gives us the following condition
\begin{equation}
f_{\cxi'\cxi}\circ\pi_\Xi(\cxi)=\pi_\Xi(\cxi')\circ f_{\cxi'\cxi},
\label{ff}
\end{equation}
while the condition following from \eqref{add-fr} does not imply any further restriction on $f_{\cxi'\cxi}$ (this is because any $f_{\cxi'\cxi}\in\Aut_q$ satisfies the condition).

It follows easily from \eqref{aut-a-g} and \eqref{xiphi} that for any $\xi\in\Xi_q$ and any $\rho\in\Aut_q$
\[
\rho^{-1}\circ\xi=\xi\circ\rho.
\]
This allows us to rewrite \eqref{ff} as
\begin{equation}
f_{\cxi'\cxi}^2=\pi_\Xi(\cxi')\circ\pi_\Xi(\cxi).
\label{f2}
\end{equation}
We see now that in order to find $f_{\cxi'\cxi}$ we have to take the square root of the r.h.s. of the equation above. Using an arbitrary but fixed $\xi^0\in\Xi_q$ we can express the equation as follows:
\[
f_{\cxi'\cxi}^2=\pi_\Xi(\cxi')\circ\xi^0\circ\xi^0\circ\pi_\Xi(\cxi)=(\pi_\Xi(\cxi')\circ\xi^0)\circ(\pi_\Xi(\cxi)\circ\xi^0)^{-1}.
\]
Since $(\pi_\Xi(\cxi)\circ\xi^0)\in\Aut_q$ one can actually find a {\em continuous} map $\mu:\check{\Xi}_q\to\check{\Aut}_q$ such that
\[
\pi_{\rm A}(\mu(\cxi))=\pi_{\Xi}(\cxi)\circ\xi^0
\]
for every $\cxi\in\check{\Xi}$. Indeed, using \eqref{aut-a-g}, \eqref{xiphi} and \eqref{R-xi-aut} to parameterize the sets ${\Xi}_q$, ${\Aut}_q$, $\check{\Xi}_q$ and $\check{\Aut}_q$ it is easy to check that if $\cxi\equiv\cxi_\beta$ and $\xi^0\equiv\xi_{e^{i\beta'}}$ for some $\beta,\beta'\in\R$ then
\[
\mu(\cxi_\beta)=\check{\rho}_{\beta'-\beta+2k\pi},
\]
where $k$ is an integer which in general may depend on $\beta$ and $\beta'$, and $2\pi$ is the period of the projection $\pi_{\rm A}$. So the map $\mu$ can be expressed as
\begin{equation}
\beta\mapsto\beta'-\beta+2k\pi.
\label{mu-beta}
\end{equation}
Clearly, the map is continuous if and only if the integer $k$ does not depend on $\beta$. Thus
\[
f_{\cxi'\cxi}^2=\pi_{\rm A}(\mu(\cxi'))\circ\pi_{\rm A}(\mu(\cxi))^{-1}=
\pi_{\rm A}(\mu(\cxi')-\mu(\cxi)).
\]
Now we can easily write down solutions of the above equation as
\begin{equation}
f_{\cxi'\cxi}=\pi_{\rm A}(\frac{\mu(\cxi')-\mu(\cxi)}{2})
\ \ \ \text{or} \ \ \
f_{\cxi'\cxi}=\pi_{\rm A}(\frac{\mu(\cxi')-\mu(\cxi)}{2}+\pi).
\label{f-result}
\end{equation}

We have to convince ourselves that the results do not depend on the choice of the map $\mu$---the map depends on $\xi^0$ ($\beta'$ in \eqref{mu-beta}) and while $\xi^0$ is fixed there are still many different choices of $\mu$ (given $\xi^0$ the integer $k$ in \eqref{mu-beta} can be chosen arbitrarily). Let us consider another map $\mu'$ given by some ${\xi^{\prime 0}}\in \Xi_q$. Then using the definition of $\mu$ and $\mu'$ we obtain
\[
\pi_{\rm A}(\mu(\cxi))=\pi_{\rm A}(\mu'(\cxi))\circ(\xi^{\prime 0}\circ\xi^0).
\]
Because $(\xi^{\prime 0}\circ\xi^0)\in\Aut_q$ there exists $\check{\rho}\in\check{\Aut}_q$ which is projected on $(\xi^{\prime 0}\circ\xi^0)$ under $\pi_{\rm A}$. Thus
\[
\mu(\cxi)=\mu'(\cxi)+\check{\rho}+2k\pi, \ \ k\in\Z.
\]
Note that since the maps $\mu$ and $\mu'$ are continuous the integer $k$ does not depend on $\cxi$. Hence we have
\[
\mu(\cxi')-\mu(\cxi)=\mu'(\cxi')-\mu'(\cxi)
\]
which means that the solutions \eqref{f-result} do not depend on the choice of $\mu$.

To complete our task we have to take into account the only nontrivial 'commutation relation' of \eqref{fr-fr} that is
\[
{\rm fr}_{e,\cxi''\cxi'}\circ{\rm fr}_{e,\cxi'\cxi}={\rm fr}_{e,\cxi''\cxi}.
\]
It is easy to see that the first solution of \eqref{f-result} satisfy the condition following from the relation above while the second one does not.

Thus we show that there exists the only continuous map
\[\label{fxixi}
\check{\Xi}^2_q\ni(\cxi,\cxi')\mapsto f_{\cxi'\cxi}\in\Aut_q
\]
which is consistent with the restrictions originating form the 'commutation relations' \eqref{sub-fr}-\eqref{fr-fr}. Consequently, the map $p^{\rm fr}$ is defined unambiguously.

\subsubsection{Conclusion}

In this way we have found all the maps corresponding to the new elementary
transformations, which allows us to define the maps $\{p_{\cg'\cg}\}$,
and finally:

\begin{df}
A natural  $\suq$ connection space  $\check{\cyl}_q$ is the
limit of the inductive family $\{\CC^{\cg},p_{\cg'\cg}\}_{\cg',\cg \in
\check{\rm Gra}}$ equipped with the $C^*$-algebra structure induced
by the structure of $\suq$.
\end{df}

Indeed,  the maps
 $p^{\rm sub},p^{\rm or},p^{\rm add}$ and $p^{\rm fr}$ are given naturally,
so the algebra $\check{\cyl}_q$ is a natural quantum connection space
built over the quantum group $\suq$.

Let us finally explain why we have not used the elements of $\Xi_q$, that is, the internal framings to define the framed graphs. If we have, then the map $p^{\rm fr}$ (see \eqref{fr-f}) would be of the following form
\[
p^{\rm fr}:=\id\ot\ldots\ot\id\ot f_{\xi'\xi},
\]
where $\xi',\xi\in\Xi_q$. In this case the relation \eqref{or-fr} would give rise to the following condition
\[
f^2_{\xi'\xi}=\xi'\circ\xi
\]
imposed on $f_{\xi'\xi}$ as a counterpart of \eqref{f2}. Using \eqref{aut-a-g} and \eqref{xiphi} to para\-met\-ri\-ze the sets ${\Xi}_q$ and ${\Aut}_q$ we convince ourselves that the above condition is equivalent to (compare with \eqref{z2ww'})
\[
z^2=\bar{w}w'
\]
as a condition imposed on $z\in U(1)$ with $w,w'\in U(1)$. Obviously one
can solve this equation for any fixed $w,w'$ but there is no
{\em continuous} map $U(1)^2\ni(w,w')\mapsto z(w,w')\ni U(1)$ such that $z(w,w')$ satisfies
the condition. Consequently, to construct an $\suq$ connection space
we would have to use $p^{\rm fr}$ discontinuous as a
function of the framing. Since there are many such maps the construction would not be natural. To avoid this we have defined framed graphs using
the elements of $\check{\Xi}_q$.

\subsection{Summary of the framing independent construction}

The set Gra of embedded graphs in $\Sigma$ is replaced by the set
$\check{\rm Gra}$ of framed
graphs. A framed graph $\check{\gamma}$ is an embedded graph
$\gamma$---the bare graph of $\check{\g}$---whose edges
are colored by elements of the covering space $\check{\Xi}_q$
to the space of internal framings $\Xi_q$,
\begin{equation}
\gamma\ni e\mapsto \check{\xi}_e\in \check{\Xi}_q.
\end{equation}
The set of framed graphs is directed by (\ref{geq}). To every framed
graph $\check{\g}$ we assign the same as before quantum group
$$\CC^{\check{\g}}:=\CC^\gamma$$
where $\g$ is the bare graph of $\check{\g}$.

 The  elementary transformations in the set Gra
 give rise to  elementary transformations in $\check{\rm Gra}$:
adding a colored edge $(e,\check{\xi}_e)$, splitting an edge
$(e=e_1\circ e_2,\check{\xi}_e)$ into
$((e_1,\check{\xi}_e),\,(e_2,\check{\xi}_e))$, reorienting
$(e,\check{\xi}_e)$ into $(e^{-1},\check{\xi}_e)$. The new elementary
transformation is changing  a coloring of $(e,\check{\xi}_e)$ into
$(e,\check{\xi}'_e)$.

The injective homomorphisms $p_{\check{\g}'\check{\g}}$ corresponding
to the first two elementary transformations are the same as in the previous
construction of $\cyl_Q(\xi)$ (see \eqref{sub-phi},
\eqref{add-in}).

The homomorphism $p_{\check{\g}'\check{\g}}$
corresponding to the edge reorienting is defined
as in the construction of $\cyl_Q(\xi)$ (see \eqref{or-xi})
with the new element: substitution $\xi=\pi_{\Xi}(\check{\xi}_e)$.

The homomorphisms
$p_{\check{\g}'\check{\g}}$ corresponding to the last transformation---to
the coloring change---is given by (\ref{fr-f}). The key technical difficulty
was  finding a natural  map $f_{\check{\xi}'\check{\xi}}$  which satisfies the condition
(\ref{ff}).

The result is a new inductive family
$(\CC^{\check{\g}},
p_{\check{\g}'\check{\g}})_{{\check{\g}'\geq\check{\g}}\in
\check{\rm Gra}}$.

Our final result is the  limit $\check{\cyl}_Q$ of the inductive family
named the framing independent $\suq$ space of connections.
The relation between the very $\check{\cyl}_Q$ and the previous
$\cyl_Q(\xi)$ is described by the following lemma:

\begin{lm}
For every $\xi\in\Xi_q$, the $\suq$ connection space
$\cyl_Q(\xi)$ is isomorphic with $\cyl_Q$.
\end{lm}

\noindent{\bf Proof.}
To construct an isomorphism we just need
to choose
$$ \check{\xi}\in \pi_\Xi^{-1}(\xi).$$
Next, we define
\[
\cyl_Q(\xi)\ni [a_\gamma]\ \mapsto\ [a_{\check{\gamma}}=a_\g]\in \cyl_Q,
\]
where $\check{\g}$ is a framed graph obtained by coloring the graph
$\g$ with the constant map
$$\g\ni e\mapsto\check{\xi}.$$
$\blacksquare$

\section{Discussion, outlook}

Where are the connections themselves? In the case of the Ashtekar-Isham $C^*$-algebra built over the classical Lie group $G$, each pure state on the algebra defines a generalized connection. Among them there are regular connections which satisfy some smoothness conditions \cite{barret}. Therefore, given a quantum group connection space $\cyl_Q(\xi)$, quantum group connections can be probably defined as pure states on $\cyl_Q(\xi)$ subjected to smoothness conditions as counterparts of those described in \cite{barret}.

Application of the Ashtekar-Isham $C^*$-algebra in LQG resulted in defining some additional structures on it e.g.:
\begin{enumerate}
\item a natural, diffeomorphism invariant state on the algebra generated by the Haar measure on $G$ \cite{al-hoop};
\item a subalgebra of smooth cylindrical functions;
\item differential operators (flux operators) on the subalgebra \cite{area};
\item quantum geometry operators like e.g. the area and volume operators \cite{area,vol}.
\end{enumerate} 
A natural question is whether these structures can be generalized to the non-commutative $C^*$-algebras constructed in this paper. The answer is positive at least in the case of the first three ones.

On every $\cyl_Q(\xi)$ built over any compact quantum group $(\CC,\Phi)$ (including the natural algebra $\check{\cyl}_q$ built over $\suq$) there exists a natural state generated by the Haar measure $h$ on the quantum group (see Definition \ref{haar-df} originally stated in \cite{slw}). Its construction \cite{oko-phd} is fully analogous to the construction of the state on the Ashtekar-Isham algebra. Given algebra $\CC^{\g}\cong\CC^N$ associated with a graph $\gamma$ of $N$ edges we define
\[
h_\gamma:=h\ot\ldots\ot h:\CC^{\g}\to \C.
\]    
Recall now that all homomorphisms $p_{\g'\g}$ are compositions of the elementary homomorphisms $p^{\rm sub},p^{\rm or}$ and $p^{\rm add}$ (and $p^{\rm fr}$ in the case of $\check{\cyl}_q$). Using Theorem \ref{haar-thr} and Lemma \ref{haar-pres} it is easy to show that the states $\{h_\g\}$ are compatible with the elementary homomorphisms and consequently with every homomorphism $p_{\g'\g}$ i.e. if $\g'\geq\g$ then 
\[
h_{\g}=(p_{\g'\g})_*h_{\g'}.
\]      
This means that the family $\{h_\g\}$ defines a functional on $\cyl_Q(\xi)$ which is the natural state.  

The subalgebra of smooth cylindrical functions can be also easily generalized to the case of any compact quantum group. The key observation is that the Hopf algebra $H$ of a quantum group can serve as a domain of suitably defined differential operators on the quantum group \cite{slw-diff}. Given an inductive family $(\CC^\g,p_{\g',\g})$ defining $\cyl_Q(\xi)$, there exists a corresponding family $(H^\g,p_{\g',\g})$, where
\[
H^\g:=H\ot\ldots\ot H\subset \CC^\g.
\]   
The inductive limit $\cyl^\infty_Q(\xi)$ of $(H^\g,p_{\g',\g})$ is a dense subalgebra of $\cyl_Q(\xi)$ and a natural counterpart of the subalgebra of smooth cylindrical functions \cite{oko-phd}. 

Applying a differential calculus defined on a quantum group \cite{slw-diff} one can define on $\cyl^\infty_Q(\xi)$ counterparts of flux operators known from LQG \cite{oko-phd}. Note however that, given a compact quantum group, there is in general no canonical differential calculus \cite{slw-diff,slw-suq}.

Generalizations of quantum geometry operators to the case of noncommutative algebras $\cyl_Q(\xi)$ to the best knowledge of the authors were not studied yet.   

\paragraph{Acknowledgments}   We would like to thank professors Abhay Ashtekar, John Baez,  Ludwik D\c{a}browski, Piotr Hajac, Shahn Majid, Wies{\l}aw Pusz, Piotr So{\l}tan and Stanis{\l}aw L. Woronowicz for discussions and helpful comments.  The work was partially supported by the Polish Ministerstwo Nauki i Szkolnictwa Wy\.zszego grant 1 P03B 075 29, by 2007-2010 research project N202 n081 32/1844 , the National Science Foun-dation (NSF) grant PHY-0456913 and by the Foundation for Polish Science grant "Master".

\appendix

\section{Terminology used in the quantum group section}
Now let us give precise meanings of some terms. Consider $*$-algebras $\cal B$ and $\cal B'$ and a linear map $\zeta: {\cal B}\rightarrow {\cal B}'$. We will say that $\zeta$ is a $*$-homomorphism of the algebras if and only if
\[
\zeta(ab)=\zeta(a)\zeta(b)\ \ \ \text{and} \ \ \ \zeta(a^*)=\zeta(a)^*
\]
for every $a,b\in {\cal B}$. A map $\zeta$ is an isomorphism of $*$-algebras $\cal B$ and $\cal B'$ if it is an invertible $*$-homomorphism between the algebras.
If ${\cal B}={\cal B}'$ then we say that an isomorphism $\zeta$ is an automorphism of a $*$-algebra $\cal B$.

If $\cal B$ and $\cal B'$ are $C^*$-algebras  and $\zeta$ is any $*$-homomorphism from $\cal B$ into $\cal B'$ then
\[
\|\zeta(a)\|'\leq \|a\|
\]
for every $a\in{\cal B}$, where $\|\cdot\|$ and $\|\cdot\|'$ are norms on $\cal B$ and $\cal B'$ respectively. If $\zeta$ is injective then $\|\zeta(a)\|'=\|a\|$.

{ Let $\CC$ be a $C^*$-algebra of a compact quantum group $(\CC,\Phi)$.
We will distinguish between automorphisms of the $*$-algebra $\CC$ and
{\em comultiplicative} automorphisms of the algebra by calling the latter
ones automorphisms of a {\em quantum group} $(\CC,\Phi)$
(for short: automorphism of a quantum group $\CC$).} Finally, an algebra $\cal B$ is unital if it possesses a unit, and a map between unital algebras $\cal B$ and $\cal B'$ is unital, if it maps the unit of $\cal B$ to the unit of $\cal B'$.

Given two $C^*$-algebras $\CC_1$ and $\CC_2$, the tensor product $\CC_1\ot\CC_2$ is defined as a completion of the algebraic tensor product $\CC_1\ot_{\rm alg}\CC_2$ with respect to a norm defined as follows: let $\pi_1$ and $\pi_2$ be any {\em faithful} (i.e. injective) $*$-representations  of $\CC_1$ and $\CC_2$, respectively, on Hilbert spaces $\h_1$ and $\h_2$. Then \cite{takesaki}
\[
\| \sum_i a_i\ot b_i\|:=\|\sum_i \pi_1(a_i)\ot\pi_2(b_i)\|',
\]
where $\|\cdot\|'$ is a norm on the algebra $B(\h_1\ot\h_2)$ (one can show that the norm on $\|\cdot\|$ does not depend on the choice of the representations). Let $\zeta_i:\CC_i\mapsto\CC'_i$ ($i=1,2$), where $\{\CC'_i\}$ are $C^*$-algebras, be $*$-homomorphisms (injective $*$-homomorphism). Then $\zeta_1\ot\zeta_2$ is $*$-homomorphism (injective $*$-homomorphism) from $\CC_1\ot\CC_2$ onto  $\CC'_1\ot\CC'_2$. If  $\chi_i:\CC_i\mapsto\C$ is  positive (faithful positive) functional on $\CC_i$ then $\chi_1\ot\chi_2$ is  positive  (faithful positive) functional on $\CC_1\ot\CC_2$ \cite{takesaki}.

\section{Automorphisms and internal framings of $\suq$ }

In the present section we are going to derive all automorphisms of $\suq$ for $q\neq -1$ and complete the partial derivation of the internal framings performed in Sections \ref{subsec:suqcon} and \ref{su2con}. This task is rather technical, so before we will start the derivation we need to introduce briefly some notions taken from \cite{ slw}, where the reader will find an exhaustive description of them.  

\subsection{Preliminaries}

\subsubsection{ Irreducible unitary representations of quantum group and its Hopf algebra}

Let $K$ be a finite-dimensional Hilbert space, and $B(K)$ the algebra of bounded operators on $K$. Consider an element $u$ of $B(K)\otimes\CC$,
\[
u=\sum_s A_s\otimes a_s,
\]
where the sum is finite. Given an orthonormal basis $(e_i)$ of $K$, we define matrix element of $u$:
\[
u_{ij}:=\sum_s \scal{e_i}{A_s e_j} a_s\in\CC.
\]

\begin{df}
We say that $u$ is a finite dimensional representation of a quantum group $(\CC,\Phi)$ on the (finite-dimensional) Hilbert space $K$ if and only if
\begin{equation}
(\id\otimes\Phi)u=u^{12}u^{13}\in B(K)\otimes\CC\otimes\CC
\label{repr}
\end{equation}
where $u^{12}:=\sum_s A_s\otimes a_s\otimes I$ and $u^{13}:=\sum_s A_s\otimes I\otimes a_s$.
\end{df}

Let $K,K'$ be finite dimensional Hilbert spaces and let $u\in B(K)\ot\CC$ and $u'\in B(K')\ot\CC$ be representations of $(\CC,\Phi)$. The representation $u$ is equivalent to $u'$ if and only if  there exists an invertible intertwining operator $W\in B(K,K')$ such that
\begin{equation}
(W\ot I)u=u'(W\ot I)\in B(K,K')\ot\CC.
\label{equiv-rep}
\end{equation}
Let $S\in B(K)$. The representation $u$ is irreducible iff
\[
(S\ot I)u=u(S\ot I)\Longrightarrow S=\lambda\,\id,
\]
for some nonzero $\lambda\in\C$.

A representation $u$ is unitary if $u$ is a unitary element of $B(K)\otimes\CC$, i.e.:
\begin{equation}
u^*u=I_{B(K)\otimes\CC}=uu^*,
\label{unit-0}
\end{equation}
where $u^*:=\sum_s A^*_s\otimes a^*_s$ and $I_{B(K)\otimes\CC}$ is the unit of the algebra ${B(K)\otimes\CC}$.

\begin{thr}[Woronowicz]
Any (not necessary finite-dimensional\footnote{A definition of an infinite-dimensional representation of a compact quantum group can be found in \cite{slw}.}) u\-nitary representation of a compact quantum group is a direct sum of finite dimensional irreducible unitary representations of the group.
\end{thr}

\begin{thr}[Woronowicz]
Let $H$ be the set of all linear combinations of matrix elements of all irreducible unitary representations of $(\CC,\Phi)$. Then $H$ is a dense $*$-subalgebra of $\CC$ and $\Phi(H)\subset H\otimes_{\rm alg} H$. Moreover, the triplet $(H,\Phi|_H, *|_H)$ is a unital Hopf $*$-algebra.
\label{hopf-thr}
\end{thr}

\subsubsection{The Haar measure}

\begin{thr}[Woronowicz]
Given compact quantum group $(\CC,\Phi)$, the\-re exists a unique state (positive normalized functional) $h$ on $\CC$ such that
\begin{equation}
(h\otimes\id)\Phi(a)=h(a)I=(\id\otimes h)\Phi(a).
\label{haar-state}
\end{equation}
(Normalization means that $h(I)=1$.)
\label{haar-thr}
\end{thr}

\begin{df}
The Haar measure on a quantum group $(\CC,\Phi)$ is the functional $h$ described by Theorem \ref{haar-thr}.
\label{haar-df}
\end{df}

\noindent {\bf Remark.} In general the Haar measure $h$ $(i)$ is {\em not} faithful state on $\CC$, $(ii)$ is {\em faithful} on its Hopf algebra $H$ and $(iii)$ its norm is equal $1$: $\|h\|=1$.

The Haar measure defines a functional $\varphi$  on the Hopf algebra $H_q$: given an irreducible unitary representations ones defines \cite{slw}:
\begin{equation}
h(u^{\a*}_{ij}u^{\a'}_{i'j'})=\frac{1}{M_\a}\delta_{\a\a'}\varphi(u^\a_{i'i})\delta_{j'j}\label{ort}
\end{equation}
where
\[
M_\a:=\varphi(\sum_k u^\a_{kk}).
\]

\begin{lm}
Suppose that $\zeta:\CC\rightarrow\CC$ is a linear (anti)comultiplicative invertible map. If $\zeta(I)=I$  then $\zeta$ preserves the Haar measure on $(\CC,\Phi)$.
\label{haar-pres}
\end{lm}

\noindent {\bf Proof.} Assume that $\zeta$ is comultiplicative.  Notice that the functional $h\circ\zeta$ is normalized, i.e. $h\circ\zeta(I)=1$. We have:
\begin{multline*}
(h\circ\zeta\ot \id)\Phi(a)=\zeta^{-1}[ \, (h\ot\id)(\zeta\ot \zeta)\Phi(a)\, ]=\\=
\zeta^{-1}[ \, (h\ot \id)\Phi(\zeta(a)) \, ]=\zeta^{-1}[\, h(\zeta(a))I\, ]=h\circ\zeta(a)I,
\end{multline*}
which by virtue of uniqueness of the Haar measure means that $h\circ\zeta=h$. The proof for an anticomultiplicative $\zeta$ is similar. $\blacksquare$

\begin{lm}
Let $H$ be the Hopf algebra of a compact quantum group $(\CC,\Phi)$ and let $\zeta$ be an invertible and (anti)comultiplicative linear map from $H$ onto itself. If $\zeta(I)=I$ then $\zeta$ preserves the Haar measure on $\CC$ restricted to $H$.
\label{hopf-haar}
\end{lm}

\noindent {\bf Proof.} The proof of this lemma coincides with the Proof of Property 4 of the Theorem 4.2 in the second paper of \cite{slw}. $\blacksquare$


\begin{lm}
Let $H'$ be a dense $*$-subalgebra of $\CC$, where $(\CC, \Phi)$ is a compact quantum group and  let $\zeta$ be an automorphism of $H'$ preserving the Haar measure restricted to $H'$. If the Haar measure is faithful on $\CC$, then  $\zeta$ is uniquely extendable to an automorphism of $\CC$.
\label{extend}
\end{lm}

\noindent{\bf Proof.} To prove\footnote{The proof follows the proof of Theorem 1.6 in the first work of \cite{slw}.} the lemma it is enough to show that for every $a\in H'$
\[
\|\zeta(a)\|=\|a\|.
\]

 Let us define a (semi-definite, in general,) scalar product on $\CC$
\begin{equation}
\scal{a}{b}:=h(a^*b).
\label{scal}
\end{equation}
The scalar product can be projected to a definite one on a quotient space $\CC/{\cal I}_h$, where
\[
{\cal I}_h:=\{a\in \CC \ | \ h(a^*a)=0 \}
\]
is a left ideal of $\CC$. Denote by $\h_\CC$ a Hilbert space obtained as a completion of $\CC/{\cal I}_h$ equipped with the projected scalar product and a norm defined by the product. Note that $\h_\CC$ is in fact the carrier Hilbert space of GNS representation of $\CC$ given by the Haar measure.

It is clear that $\zeta$ preserves the scalar product \eqref{scal} restricted to $H'$. Because the norm of the Haar measure $h$ is equal $1$ we have for every $a\in\CC$
\[
\|a\|^{\prime 2}=|h(a^*a)|\leq\|a^*a\|=\|a\|^2,
\]
where the last equation holds by virtue of the definition of $C^*$-algebra, and $\|\cdot\|'$ is the norm on $\h_\CC$. Thus the density of $H'$ in $\CC$ implies the density of $H'/{\cal I}_h$ in $\h_\CC$. This fact together with  $\zeta(H')=H'$ mean that $\zeta$  can be uniquely extended to a unitary map $U_\zeta$ on $\h_\CC$. Recall that  $\h_\CC$ is the carrier space of the GNS representation $\pi$ of $\CC$ given by the Haar measure. By a direct calculation we obtain
\[
\pi(\zeta(a))=U_\zeta\pi(a)U_\zeta^*
\]
for every $a\in H'$. We assumed that the Haar measure is faithful on $\CC$. Consequently  the representation $\pi$ is an injective $*$-homomorphism from $\CC$ into $B(\h_{\CC})$ (i.e. into the $C^*$-algebra  of bounded operators on $\h_\CC)$. Every such homomorphism preserves the norm, thus for every $a\in H$
\[
\|\zeta(a)\|=\|\pi(\zeta(a))\|_B=\|U_\zeta\pi(a)U_\zeta^*\|_B=\|\pi(a)\|_B=\|a\|,
\]
where $\|\cdot\|_B$ is the norm on $B(\h_{\CC})$. $\blacksquare$

\subsection{Fundamental representation of $\suq$}

In the case of $\suq$ for $q\neq 1$ there exists a unique two-dimensional irreducible unitary representation of the quantum group \cite{slw-suq}. We will call it a {\em fundamental representation}. There exists an orthonormal basis of the carrier Hilbert space of the representation such that the matrix of the representation defined by the basis is of the form \eqref{su-fund}. Thus matrix elements of fundamental representation are generators of the Hopf algebra $H_q$ of $\suq$ (see Section \ref{subsec:suq}). To define a $*$-algebra automorphism of $H_q$ it is enough to define its action on the generators, but it does not mean that it can be automatically extended to a $C^*$-algebra automorphism of $\CC_q$. Since we are going to derive some automorphism of $\CC_q$ we need the following lemma:

\begin{lm}
Let $(u_{ij})$ be a matrix of the fundamental representation of $\suq$. Assume that there is a linear map
\[
\zeta:\ {\rm span}\{u_{ij},I\}\rightarrow{\rm span}\{u_{ij},I\}
\]
such that
\begin{enumerate}
\item it preserves ${\rm span}\{u_{ij}\}$ and maps $I$ to $I$;
\item it preserves $*$-involution on ${\rm span}\{u_{ij},I\}$;
\item $\zeta\equiv\zeta^1$  is invertible with the inverse map $\zeta^{-1}$;
\item
\begin{equation}
\sum_k \zeta^{\pm1}(u^*_{ki})\zeta^{\pm1}(u_{kj})=\delta_{ij}I=\sum_{k} \zeta^{\pm1}(u_{ik})\zeta^{\pm1}(u^*_{jk}),
\label{suq-1}
\end{equation}
\item either
\begin{equation}
\Phi(\zeta(u_{ij}))=(\zeta\ot\zeta)\Phi(u_{ij})
\label{zeta-com}
\end{equation}
or
\begin{equation}
\Phi(\zeta(u_{ij}))=\sigma(\zeta\ot\zeta)\Phi(u_{ij}).
\label{zeta-acom}
\end{equation}
\end{enumerate}
Then there exists a unique automorphism $\tl{\zeta}$ of the $C^*$-algebra $\CC_q$ such that it coincides with $\zeta$ if restricted to ${\rm span}\{u_{ij},I\}$. If $\zeta$ satisfies \eqref{zeta-com} (\eqref{zeta-acom})then $\tl{\zeta}$ is comultiplicative (anticomultiplicative).
\label{extension}
\end{lm}

\noindent {\bf Proof.} The elements $\{u_{ij}\}$ subjected to commutation relations \eqref{unit-0}\footnote{It is not difficult to check that in the case of $\suq$ \eqref{unit-0} is  equivalent to the commutation relations \eqref{suq} defining the Hopf $*$-algebra $H_q$.} are generators of the $*$-algebra $H_q$. Therefore the first four assumptions guarantee that $\zeta$ can be unambiguously extended to an invertible $*$-preserving endomorphism of the $*$-algebra $H_q$. The endomorphism is either comultiplicative or anticomultiplicative if either \eqref{zeta-com} or \eqref{zeta-acom} is satisfied. Now Lemma \ref{hopf-haar} ensures that the endomorphism preserves the Haar measure on $\suq$ restricted to $H_q$. This means that we can apply Lemma \ref{extend} to conclude that the endomorphism can be extended in a unique way to an automorphism of $\CC_q$ (the Haar measure is faithful on $\suq$ \cite{slw}) which will be either comultiplicative or anticomultiplicative. $\blacksquare$

\subsection{Automorphisms of $\suq$ \label{ap-auth}}

Now we are ready to start the derivation of all automorphisms of $\suq$. Recall that an automorphism of a compact quantum group $(\CC,\Phi)$ is a comultiplicative automorphism of $C^*$-algebra $\CC$ (notice also that such a map preserves necessary the unit of the algebra). In the sequel the symbol $u$ will denote the fundamental representation of $\suq$, while $(u_{ij})$ denotes the matrix  of the representation given by an orthonormal basis of the carrier Hilbert space.

The following lemma describes the necessary condition for a linear map from $\CC_q$ onto itself to be an automorphism of $\suq$.

\begin{lm}
If $\rho$ is an automorphism of $\suq$ then its action on the generators $\a$ and $\g$ is the following:
\begin{equation}
\begin{aligned}
\rho(\a)\equiv\rho_{z,x}(\alpha)&=z\bar{z}\a-\bar{z}\bar{x}\g+zx\g^*+x\bar{x}\a^*\\
\rho(\g)\equiv\rho_{z,x}(\g) & =\bar{z}x\a+\bar{z}^2\g+x^2\g^*-x\bar{z}\a^*.
\end{aligned},
\label{aut-q-1}
\end{equation}
 where $z\bar{z}+x\bar{x}=1$. If $q^2\neq 1$ then in the above formulas $x=0$. The equality $\rho_{z,x}(\cdot)=\rho_{z',x'}(\cdot)$ holds if and only if $z'=\pm z$ and $x'=\pm x$.

Equivalently, given $(u_{ij})$ of the form \eqref{su-fund},
\begin{equation}
\rho(u_{ij})=\sum_{kl}S_{ik}u_{kl}S^{-1}_{lj},
\label{aut-sus}
\end{equation}
where $(S_{ij})$ is an $SU(2)$-matrix, which is required to be diagonal if $q^2\neq1$.

\label{nec-auth}
\end{lm}

\noindent {\bf Proof.} Let $u$ be the fundamental representation of $\suq$ with the carrier Hilbert space $K$, and let $\rho$ be an automorphism of the quantum group. Then
\[
u':=(\id\ot\rho)u\in B(K)\ot\CC_q
\]
is a two-dimensional unitary irreducible representation of $\suq$. The representation theory of $\suq$ \cite{slw-suq} ensures us that $u'$ is equivalent to $u$, thus there exists an invertible map $S\in B(K)$ such that
\[
(S\ot I)u=u'(S\ot I)
\]
or equivalently
\begin{equation}
\rho(u_{ij})=u'_{ij}=\sum_{kl}S_{ik}u_{kl}S^{-1}_{lj}.
\label{aut-S}
\end{equation}
Notice that if two maps $S,S'\in B(K)$ satisfy the above condition  then there exists  nonzero $\lambda\in\C$ such that
\begin{equation}
S=\lambda S'.
\label{SS'}
\end{equation}
By virtue of Lemma \ref{haar-pres} $\rho$ preserves the Haar measure on $\suq$. Because $\rho$ is multiplicative it preserves as well the scalar product (\ref{scal}) on $\h_{\CC_q}$, thus in particular
\[
\scal{\rho(u_{ij})}{\rho(u_{i'j'})}=\scal{u_{ij}}{u_{i'j'}}.
\]
By means of (\ref{ort}) we obtain from the latter equation:
\[
(\sum_{k'k}S_{i'k'}\varphi(u_{k'k})\overline{S}_{ik})\ \delta_{n'n}\ = \ \varphi(u_{i'i})\ (\sum_{j}\overline{S}_{jn}S_{jn'}).
\]
The solution of the above equation is of the form
\begin{gather}
\sum_{k'k}S_{i'k'}\varphi(u_{k'k})\overline{S}_{ik}=\lambda \varphi(u_{i'i}),\label{S-f-1}\\
\sum_{j}\overline{S}_{jn}S_{jn'}=\lambda\delta_{n'n}\label{S-delta},
\end{gather}
for some nonzero $\lambda\in\C$. Consider first the second of the latter equations. The trace of it implies that $\lambda$ is real and positive. The equation implies as well that $\det S=\lambda e^{i\phi}$ for some $\phi\in\R$. Define:
\begin{equation}
S'_{ij}:=\frac{e^{-\frac{i}{2}\phi}}{\sqrt{\lambda}}S_{ij}.
\label{det-S}
\end{equation}
Then $\det S'=1$ and (\ref{S-delta}) can be expressed as
\begin{equation}
\sum_{j}\overline{S'}_{jn}S'_{jn'}=\delta_{n'n}.
\label{S'-delta}
\end{equation}
We conclude that the action \eqref{aut-S} of $\rho$ on matrix elements of the fundamental representation can be described by means of an $SU(2)$-matrix.

Recall that every automorphism has to preserve the $*$ involution. The fundamental representation $u$ of $\suq$ is equivalent to its complex conjugate\footnote{Denote by $K'$ the dual Hilbert space to $K$ (at this moment we do not identify the two spaces as it is usually done). We define complex conjugate to a representation $w$ of a quantum group on $K$ as \cite{slw}
\[
\bar{w}:=((*\circ t)\ot*)w=\sum_s (A^t_s)^*\ot a^*_s\in B(K')\ot\CC,
\]
where $t$ is the transposition map from $B(K)$ to $B(K')$.} $\bar{u}$ \cite{slw}, i.e. there exists an invertible matrix $E_{ij}$ such that
\begin{equation}
u^*_{ij}=\sum_{kl}E^{-1}_{ik}u_{kl}E_{lj}.
\label{u-star-E}
\end{equation}
The condition $\rho(u^*_{ij})=\rho(u_{ij})^*$ can be expressed as
\[
\sum_{klnm}E_{ik}\overline{S'}^{-1}_{kl}E_{ln}^{-1}S'_{nm}u_{mj}=\sum_{klnm}u_{im}E_{mk}\overline{S'}^{-1}_{kl}E_{ln}^{-1}S'_{nj}.
\]
Because $u$ is irreducible there exists nonzero $\lambda_1\in \C$ such that
\begin{equation}
\sum_{kln}E_{ik}\overline{S'}^{-1}_{kl}E_{ln}^{-1}S'_{nm}=\lambda_1\delta_{im}.
\label{ES}
\end{equation}

If $(u_{ij})$ is of the form (\ref{su-fund}) then:
\[
E^{-1}_{ij}=
\begin{pmatrix}
0 & q \\
-1 & 0
\end{pmatrix}\ \ \text{and}\ \ \varphi(u_{ij})=
\begin{pmatrix}
|q| & 0 \\
0 & |q|^{-1}
\end{pmatrix}.
\]
Denoting $S'_{11}=z=\overline{S'}_{22}$ and $S'_{21}=x=-\overline{S'}_{12}$ we get from (\ref{S-f-1}) and (\ref{ES}) a system of equations
\begin{equation}
\begin{gathered}
z\bar{z}|q|+x\bar{x}|q|^{-1}=|q|, \ z\bar{x}(|q|-|q|^{-1})=0, \ z\bar{z}|q|^{-1}+x\bar{x}|q|=|q|^{-1},\\
qx=\lambda_1 x, \ z=\lambda_1 z, \ \lambda_1^2=1.
\end{gathered}
\label{system}
\end{equation}

Solutions of the system are the following:
\begin{enumerate}
\item for $q=1$ we do not get any restrictions on $SU(2)$ matrix $(S'_{ij})$, and $\lambda_1=1$,
\item for $q^2\neq 1$ we get
\[
z\bar{z}=1, \ x=0, \ \lambda_1=1.
\]
\end{enumerate}
$\blacksquare$

The following lemma completes the description of ${\rm Aut}_q$:

\begin{lm}
Let $\rho$ be a linear map from ${\rm span}\{u_{ij},I\}\subset\CC_q$ into itself, where $(u_{ij})$ is of the form \eqref{su-fund}. Assume moreover that $\rho(I)=I$ and \eqref{aut-sus} is satisfied for either
\begin{enumerate}
\item any $SU(2)$ matrix $(S_{ij})$ if $q=1$;
\item any diagonal $SU(2)$ matrix $(S_{ij})$ if $q^2\neq 1$.
\end{enumerate}
Then $\rho$ can be unambiguously extended to an automorphism of $\suq$.
\end{lm}

\noindent {\bf Proof.} It is easy to check that the map under consideration satisfies all assumptions of Lemma \ref{extension}. $\blacksquare$

\subsection{Internal framings of  $\suq$ \label{app-anti}}

\subsubsection{Internal framings of $SU_1(2)$ }

To complete the derivation of all internal framings in the case of $SU_1(2)$ it is enough to show that the explicite formulae \eqref{xi-zx} follow from \eqref{Xpi}. Note first that the generators $\alpha$ and $\gamma$ can be now considered as functions on $G=SU(2)$. Thus given $g\in SU(2)$ we can write 
\[
(\xi_X(\alpha))(g)=((R^*_{X,\pi}\circ\kappa)(\alpha))(g)=\alpha^*(R_{X,\pi}(g)),
\]    
since $\kappa(\alpha)=\alpha^*$. It is commonly known that the action of the automorphism $R_{X,\pi}$ of $SU(2)$ corresponding to a rotation around the axis $X$ for the angle $\pi$ can be expressed as
\[
R_{X,\pi}(g)=\sigma g\sigma^{-1},
\]  
where $\sigma$ an $SU(2)$-matrix of the form
\[
(\sigma_{ij})=i
\begin{pmatrix}
x & \bar{z}\\
z & -x
\end{pmatrix}
\]
with $z\in\C$ and $x\in \R$ satisfying $z\bar{z}+x^2=1$ (the axis in given by $X=({\rm Re}z,{\rm Im}z,x)$). Taking into account that $\alpha^*(g)=g_{22}$ we obtain the first formula of \eqref{xi-zx} by a direct calculation. The second one is derived analogously.     

\subsubsection{Internal framings of proper $\suq$  }  

If $q^2\neq 1$ then the antipode $\kappa$ does not satisfy the definition of an internal framing \cite{slw}. As a hint for guessing an example $\xi_0$ of an internal framing let us use the following observation: according to Lemma \ref{nec-auth} every automorphism of the proper $\suq$ preserves the generator $\alpha$. Assume then that it is also true in the case of internal framings of $\suq$ and set $x=0$ in the formulae \eqref{xi-zx} describing the internal framings of $SU_1(2)$. For simplicity let us also set $z=1$. Thus we obtain (see \eqref{xi-ex}):
\[
\xi_0(\alpha)=\alpha, \ \ \ \xi_0(\gamma)=\gamma^*.
\]      
Assuming that $\xi_0$  preserves the unit $I$ and the $*$-involution it is easy to check that $\xi_0$ extended to a linear map from ${\rm span}\{u_{ij},I\}$ onto itself satisfies the assumptions of Lemma \ref{extension} with \eqref{zeta-acom}. Hence $\xi_0$ can be extended to an anticomultiplicative automorphism of $\CC_q$. It is also easy to check that the automorphism is involutive.


\end{document}